\documentclass[%
 aip,
 amsmath,amssymb,
reprint,%
]{revtex4-1}
\usepackage[pdftex]{graphicx}

\usepackage{dcolumn}
\usepackage{bm}
\usepackage{tcolorbox}
\usepackage[utf8]{inputenc}
\usepackage{etoolbox}
\usepackage{siunitx,physics,float,mathtools}
\usepackage{empheq}
\usepackage{here,overpic}
\usepackage{xcolor}
\makeatletter
\def\@email#1#2{%
 \endgroup
 \patchcmd{\titleblock@produce}
  {\frontmatter@RRAPformat}
  {\frontmatter@RRAPformat{\produce@RRAP{*#1\href{mailto:#2}{#2}}}\frontmatter@RRAPformat}
  {}{}
}%
\makeatother
\begin{document}


\title{Nonmonotonic concentration dependence of the self-diffusion coefficient of surfactants in wormlike micellar solutions}
\author{Yusuke Koide}
\author{Takato Ishida}%
\author{Takashi Uneyama}%
\author{Yuichi Masubuchi}%
 \email{koide.yusuke.k1@f.mail.nagoya-u.ac.jp}
\affiliation{ 
 Department of Materials Physics, Graduate School of Engineering, Nagoya University, Furo-cho, Chikusa, Nagoya, Aichi 464-8603, Japan
}%

\date{\today}

\begin{abstract}
We investigate the concentration dependence of surfactant diffusion in wormlike micellar solutions using dissipative particle dynamics simulations.
The simulations show that the self-diffusion coefficient of surfactants exhibits a nonmonotonic dependence on the surfactant concentration, as observed in previous experiments.
We quantitatively reveal that this nonmonotonic behavior results from the competition between micellar center-of-mass diffusion and surfactant diffusion within micelles by decomposing the mean-square displacement of surfactants into the corresponding contributions.
Furthermore, our detailed analyses demonstrate how the competition between the two diffusion mechanisms is governed by the aggregation number distribution, the dynamics of individual surfactants and micelles, and the kinetics of micellar scission and recombination.
\end{abstract}

\maketitle

\section{\label{sec:Intro}Introduction}

Surfactants consist of hydrophilic and hydrophobic components and self-assemble into micelles above a critical concentration.~\cite{Israelachvili2011-dh}
Micelles exhibit various morphologies, including spherical, rodlike, and wormlike structures, depending on the surfactant type and the environmental conditions.
Among these, wormlike micelles have chain-like structures resembling linear polymers.
As a result, wormlike micelles can impart pronounced viscoelasticity to solutions.~\cite{Yang2002-ob,Cates2006-cy,Dreiss2007-jf}
Despite their structural similarity, certain dynamical properties of wormlike micelles are distinctly different from those of polymers due to the reversible scission and recombination of micelles.
For instance, the linear viscoelasticity of concentrated wormlike micellar solutions is well described by the Maxwell model due to the relaxation process associated with micellar scission and recombination.~\cite{Cates1987-tv,Shikata1987-ct}
Another notable example is the self-diffusion coefficient of surfactants in micellar solutions, which exhibits a nonmonotonic dependence on the surfactant concentration.~\cite{Nilsson1983-sb,Ott1990-pt,Khatory1993-fj,Kato1993-yv,Kato1994-jt,Morie1995-iq,Narayanan1997-bn,Narayanan1998-wi}
This phenomenon cannot be explained by polymer theory.~\cite{doi1988theory}
As highlighted by these examples, understanding the role of scission and recombination in micellar solutions remains an important area of research.

To date, several studies have investigated the self-diffusion coefficient $D$ of surfactants in micellar solutions to characterize the dynamical behavior of micelles.
In particular, the intriguing nonmonotonic dependence of $D$ on the surfactant concentration has attracted attention.~\cite{Nilsson1983-sb,Ott1990-pt,Khatory1993-fj,Kato1993-yv,Kato1994-jt,Morie1995-iq,Narayanan1997-bn,Narayanan1998-wi}
Nilsson \textit{et al.}~\cite{Nilsson1983-sb} reported that $D$ of nonionic surfactants exhibited a nonmonotonic dependence on the concentration by using pulsed-gradient spin echo~(PGSE).
Mori\'{e} \textit{et al.}~\cite{Morie1995-iq} measured $D$ of cetylpyridinium chlorate~($\mathrm{CPClO_3}$), an ionic surfactant, with fluorescence recovery after photobleaching~(FRAP).
They found that above the entanglement concentration, $D$ decreased to a minimum value, while a further increase in the concentration led to a subsequent rise in $D$.
Ott \textit{et al.}~\cite{Ott1990-pt} investigated $D$ in reverse wormlike micellar solutions of the lecithin--isooctane--water system using FRAP.
They observed that $D$ increased with lecithin concentration when the water-to-lecithin molar ratio was above a certain value.

Several models have been proposed to explain the diffusion behavior of surfactants in wormlike micellar solutions.
Cates~\cite{Cates1987-tv} proposed a model that combines scission and recombination with reptation dynamics.
Turner \textit{et al.}~\cite{Turner1993-nb} modified this theory by considering ``bond interchange" and ``end interchange" processes.
In these models, $D$ is a monotonically decreasing function of the concentration, whereas experiments showed that $D$ increased at high concentrations.
This inconsistency between theories and experiments originates from the lack of surfactant diffusion within micelles in theoretical models.
To incorporate the contribution of surfactant molecules themselves, Jonstr{\"o}emer \textit{et al.}~\cite{Jonstroemer1991-tq} adopted the cell-diffusion model.~\cite{jonsson1986self}
Kato \textit{et al.}~\cite{Kato1993-yv} assumed that surfactants diffuse within a micelle along its contour and migrate to adjacent micelles with a certain timescale $\tau_\mathrm{mig}$.
In the high concentration regime, they neglected the contribution of the micellar center-of-mass diffusion and proposed that $D$ can be written as $D\simeq D_\mathrm{s}\propto \langle d^2\rangle /\tau_\mathrm{mig}$, where $D_\mathrm{s}$ represents the diffusion coefficient of surfactants based on the intra-micellar diffusion and inter-micellar migration, and $\langle d^2\rangle$ is the mean-square distance between the center of mass of two adjacent micelles.
Since $\tau_\mathrm{mig}$ decreases faster than $\langle d^2\rangle$ with increasing the concentration, this model successfully reproduced the increasing behavior of $D$ in the high concentration regime.
Later, Kato \textit{et al.}~\cite{Kato1994-jt} extended the theory to explain the whole concentration range by expressing $D$ as $D = D_\mathrm{m}+D_\mathrm{s}$, where $D_\mathrm{m}$ is the self-diffusion coefficient of micelles.
Similarly, Schmitt and Lequeux~\cite{Schmitt1998-pa} reported that the competition between the micellar diffusion and the surfactant diffusion on micelles is responsible for the nonmonotonicity of $D$ by relying on experimental data and phenomenological models.
Although these phenomenological models successfully explained the nonmonotonicity of $D$, their assumptions have yet to be directly confirmed, particularly regarding the detailed dynamics of surfactants within micelles and the kinetics of micellar scission and recombination. 
While molecular simulations offer a promising approach for analyzing the detailed behavior of individual molecules, no previous study has systematically examined the concentration dependence of $D$ in micellar solutions.

This study aims to directly reveal the physical mechanism underlying the nonmonotonic dependence of $D$ on the surfactant concentration using molecular simulations.
Since the enhanced diffusivity at high concentrations is attributed to the surfactant diffusion within micelles and micellar kinetics,~\cite{Jonstroemer1991-tq,Kato1993-yv,Kato1994-jt,Schmitt1998-pa} it is necessary to investigate micelles composed of surfactants over a sufficiently long timescale to observe a number of scission and recombination events.
For this purpose, we adopt the dissipative particle dynamics~(DPD) method.~\cite{Hoogerbrugge1992-ng,Espanol1995-mx}
In DPD simulations, surfactants and water are explicitly considered with a relatively low computational cost due to coarse-graining and soft repulsive interactions.
Using the DPD method, previous studies have examined various aspects of surfactant micelles, including their morphology, dynamics, kinetics, and rheology.~\cite{Yamamoto2005-sl,Arai2007-gs,Vishnyakov2013-ge,Kobayashi2019-co,Koide2022-bp,Koide2023-yb,Hendrikse2023-im}
In the present study, DPD simulations demonstrate that $D$ varies nonmonotonically with the surfactant concentration at various temperatures, consistent with experimental observations.
To explore the origin of the nonmonotonic behavior of $D$, we decompose the mean-square displacement of surfactants into contributions from micellar center-of-mass diffusion and surfactant diffusion within micelles.
A key aspect of this study is the systematic investigation of the dynamical properties of individual surfactants and micelles, including micellar center-of-mass diffusion, surfactant diffusion within a given micelle, and the lifetime and recombination time of micelles.
These detailed statistics allow us to elucidate the physical mechanism underlying the nonmonotonic variation of $D$.

\section{\label{sec:Method}METHODS}
In this study, we perform molecular simulations of nonionic surfactant solutions using the DPD method.~\cite{Hoogerbrugge1992-ng,Espanol1995-mx}
A single DPD particle represents a group of atoms and molecules.
Each DPD particle obeys Newton's laws of motion.
There are four types of forces in our DPD simulations: repulsive, dissipative, random, and bonded.
The surfactant model consists of a hydrophilic head particle and two hydrophobic tail particles, which are connected by the bond force $\bm{F}_{ij}^\mathrm{B}$ expressed as
\begin{equation}
    \bm{F}_{ij}^\mathrm{B} = -k_s (|\bm{r}_{ij}|-r_\mathrm{eq}) \bm{e}_{ij}, \label{eq:bond_force} 
\end{equation}
where $k_s$ is the spring constant, $r_\mathrm{eq}$ is the equilibrium bond distance, $\bm{r}_{ij}=\bm{r}_i-\bm{r}_j$, and $\bm{e}_{ij}=\bm{r}_{ij}/|\bm{r}_{ij}|$ with $\bm{r}_i$ being the position of the $i$-th particle.
Further details of the DPD simulations can be found in previous publications.~\cite{Koide2022-bp,Koide2023-ao}
In the following, all quantities are nondimensionalized by $k_BT_0$, $m$, and $r_c$, where $k_B$ is the Boltzmann constant, $T_0$ is the reference temperature, $m$ is the mass of the DPD particle, and $r_c$ is the cutoff distance.

\begin{figure*}
    \centering
    \begin{overpic}[width=0.8\linewidth]{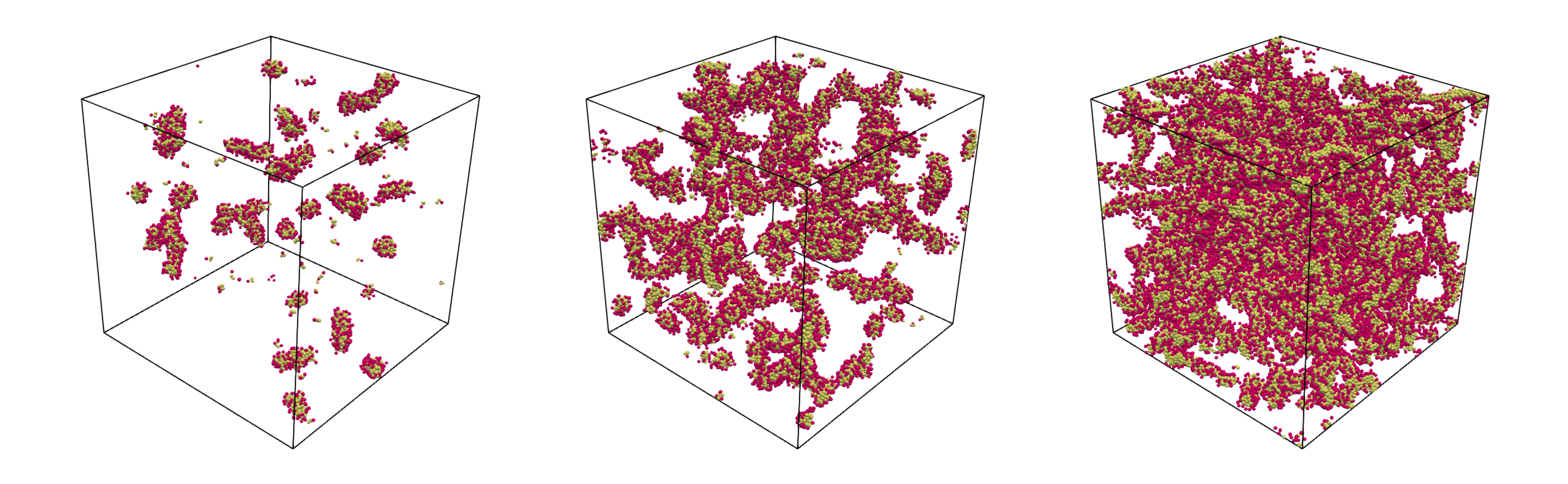} 
        \put(3,27){(a)}
        \put(35,27){(b)}
        \put(67,27){(c)}

    \end{overpic}
    \caption{Visualization of surfactant solutions for (a) $\phi=0.01$, (b) $0.05$, and (c) $0.15$. Hydrophilic and hydrophobic particles are indicated in red and yellow, respectively. For clarity, water particles are omitted.}
    \label{fig:snapshot}
\end{figure*}

In this paper, the parameters of surfactant solutions are chosen as follows:
the total number of DPD particles is $N=648\,000$; the number density of DPD particles is $\rho=3$; the random force coefficient is $\sigma=3$; the spring constant is $k_s=50$; the equilibrium bond distance is $r_\mathrm{eq}=0.8$; the repulsive force coefficients between different types of particles are $a_\mathrm{hh}=25$, $a_\mathrm{ht}=60$, $a_\mathrm{hw}=20$, $a_\mathrm{tt}=25$, $a_\mathrm{tw}=60$, and $a_\mathrm{ww}=25$, where h, t, and w denote head, tail, and water particles, respectively.
We determine the values of the repulsive force coefficient following Ref.~\onlinecite{Li2019-wy}, because wormlike micelles are formed for sufficiently large volume fractions $\phi$.
Figure~\ref{fig:snapshot} shows snapshots of surfactant solutions for $\phi=0.01$, $0.05$, and $0.15$ with $k_BT=1$.
Although most micelles are spherical or rodlike at $\phi=0.01$, the lowest value considered in this study, an increase in $\phi$ promotes surfactant aggregation, leading to the formation of wormlike micelles, as quantitatively demonstrated below~(see Fig.~\ref{fig:nag_dis}).
Note that, due to the soft-core potential used in the DPD method, there is no entanglement effect in the considered system, as previously reported in a study on polymers.~\cite{Pan2002-oq}
We implement a systematic parameter survey by changing the temperature $T$ and surfactant volume fraction $\phi$.
When varying $T$, the dissipative force coefficient $\gamma$ changes to satisfy the fluctuation-dissipation relation $\sigma^2=2\gamma k_BT$.
We have confirmed that the results remain almost unchanged when varying $\sigma$ instead of $\gamma$ within the considered range of $T$. 

We adopt the modified velocity Verlet method~\cite{Groot1997-je} as the time integration method.
Here, the parameter $\lambda$ introduced in this scheme and the time step $\Delta t$ are set to $\lambda=0.65$ and $\Delta t=0.04$, respectively.
These parameters result in sufficiently accurate temperature control.
DPD simulations are conducted in a cubic simulation box of $60\times 60\times 60$ with periodic boundary conditions.
We impose a random initial configuration and perform simulations for $12\,000\text{--}40\,000$ time units depending on the parameters, achieving statistically steady values of the potential energy and the number of micelles.
All the analyses presented in this study are done after this initial equilibration.
We perform all the DPD simulations using our in-house code.

We define a micelle using a method employed in previous studies.~\cite{Vishnyakov2013-ge,Lee2016-cx}
In this method, we judge that two surfactant molecules belong to the same cluster if a hydrophobic particle of one surfactant molecule is within $r_c(=1)$ of a hydrophobic particle of the other.
If a cluster has an aggregation number $N_\mathrm{ag}$ larger than a threshold value $n_\mathrm{mic}(=10)$, we regard the cluster as a micelle.
Micellar structures vary with the aggregation number $N_\mathrm{ag}$: micelles are spherical for $N_\mathrm{ag}\lesssim 50$, rodlike for $50\lesssim N_\mathrm{ag}\lesssim 200$, and wormlike or branched for $N_\mathrm{ag}\gtrsim 200$ in the considered systems~(see Fig.~12 in Ref.\onlinecite{Koide2022-bp}).

\section{\label{sec:result}Results}
\begin{figure}
    \centering
    \begin{overpic}[width=1\linewidth]{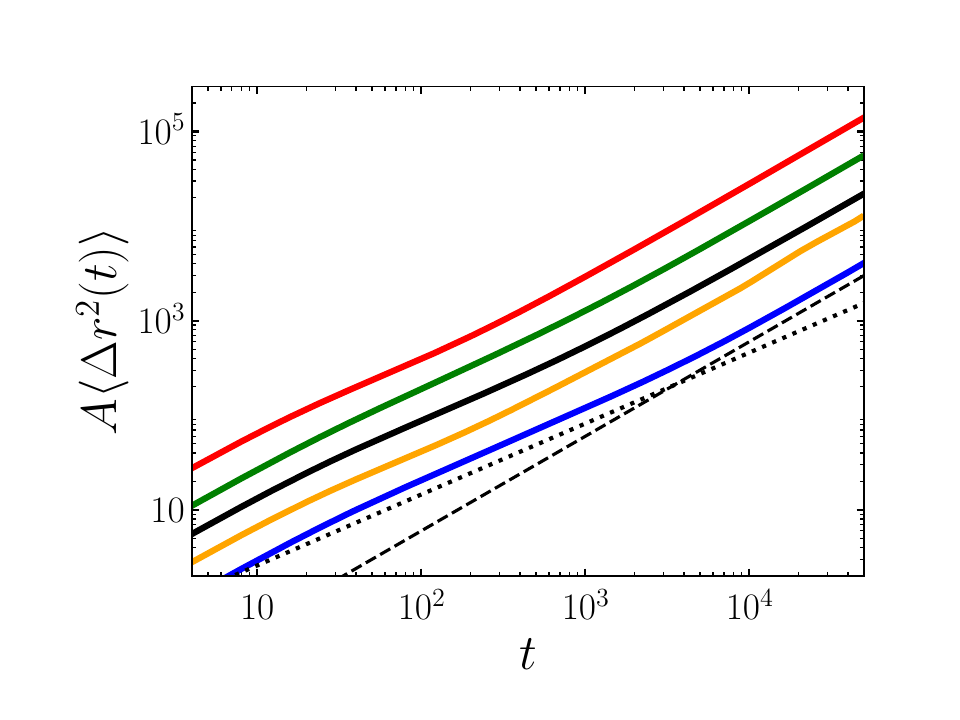} 
    \end{overpic}
    \caption{Mean-square displacement $\langle \Delta\bm{r}^2(t)\rangle$ of the center of mass of surfactant molecules for $(k_BT,\phi)=(0.9,0.05)$~(blue), $(1,0.01)$~(orange), $(1,0.05)$~(black), $(1,0.15)$~(green), and $(1.2,0.05)$~(red), being vertically shifted by $A=1$, $2$, $4$, $8$, and $16$, respectively. 
    The dotted and dashed lines indicate $\langle \Delta\bm{r}^2(t)\rangle\propto t^{0.75}$ and $\langle \Delta\bm{r}^2(t)\rangle\propto t$, respectively. 
    }
    \label{fig:msd}
\end{figure}

We evaluate the mean-square displacement~(MSD) $\langle \Delta \bm{r}^2(t)\rangle$ of surfactant molecules to investigate their diffusion behavior in micellar solutions.
Here, $\Delta \bm{r}(t)=\bm{r}(t)-\bm{r}(0)$ with $\bm{r}(t)$ being the center of mass of surfactant molecules, and $\langle\cdot\rangle$ denotes the ensemble average. 
Figure~\ref{fig:msd} shows $\langle \Delta \bm{r}^2(t)\rangle$ for various surfactant volume fractions $\phi$ and $k_BT$.
Note that $\langle\Delta \bm{r}^2(t)\rangle$ is vertically shifted by the amount $A$ for clarity. 
As reported in previous studies~\cite{Koide2022-bp,Koide2023-yb}, $\langle \Delta \bm{r}^2(t)\rangle$ exhibits a crossover from subdiffusion, where $\langle \Delta \bm{r}^2(t)\rangle\propto t^{\alpha}$ with $\alpha\simeq 0.75(<1)$, to normal diffusion, where $\langle \Delta \bm{r}^2(t)\rangle\propto t$, around the longest relaxation time of micelles.
In our systems, the subdiffusion exponent $\alpha(\simeq 0.75)$ is larger than $1/2$ predicted by the Rouse model~\cite{Rouse1953-hp} and slightly larger than $2/3$ predicted by the Zimm model.~\cite{Zimm1956-gt}
This trend is consistent with the MSD of central monomers within polymer chains reported in a previous study using DPD simulations.~\cite{Jiang2007-rf}
However, it is worth noting that, unlike in polymer systems, $\langle\Delta \bm{r}^2(t)\rangle$ (and consequently, the exponent $\alpha$) reflects the combined contributions of surfactants in micelles of various sizes as well as the kinetics of micellar scission and recombination.

\begin{figure}
    \centering
    \begin{overpic}[width=1\linewidth]{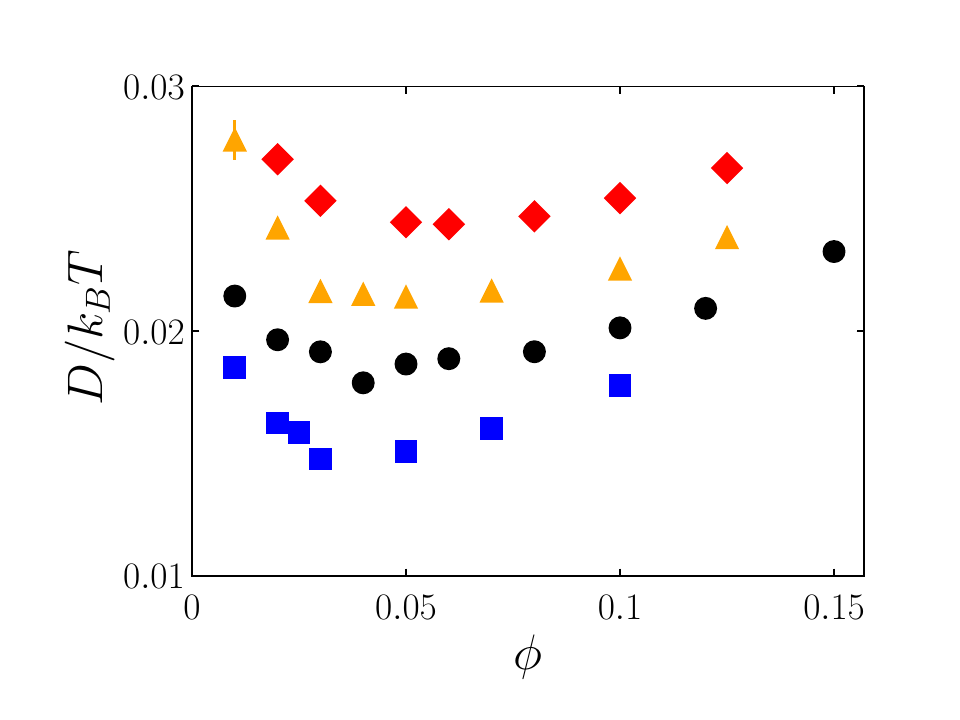} 
    \end{overpic}
    \caption{Self-diffusion coefficient $D$ of surfactant molecules normalized by $k_BT$ as a function of the volume fraction $\phi$ of surfactants for $k_BT=0.9$~(blue square), $1$~(black circle), $1.1$~(orange triangle), and $1.2$~(red diamond). The error bars denote the standard deviations for three independent simulations.}
    \label{fig:d_sur}
\end{figure}

To systematically compare the diffusivity in different systems, we evaluate the self-diffusion coefficient $D$ of surfactant molecules by fitting $\langle \Delta \bm{r}^2(t)\rangle$ to $6D t$ in the normal diffusion regime. 
Figure~\ref{fig:d_sur} shows $D$ normalized by $k_BT$ as a function of $\phi$ for various $k_BT$.
As $\phi$ increases, $D$ initially decreases and subsequently increases after reaching a minimum.
A similar nonmonotonic variation of $D$ with the surfactant concentration has been experimentally reported in previous studies.~\cite{Nilsson1983-sb,Ott1990-pt,Khatory1993-fj,Kato1993-yv,Kato1994-jt,Morie1995-iq,Narayanan1997-bn,Narayanan1998-wi}
We qualitatively reproduce the nontrivial behavior of $D$ observed in experiments with coarse-grained molecular simulations, although $D$ does not show a drastic change, probably due to the lack of entanglement in our systems.
We have confirmed that the system size is not an essential factor contributing to the nonmonotonicity of $D$ because the nonmonotonic behavior is observed for different system sizes~(see Appendix A). 
Figure~\ref{fig:d_sur} also demonstrates that this nonmonotonic dependence of $D$ emerges for different values of $k_BT$.
Incidentally, $D/k_BT$ still exhibits $k_BT$ dependence at fixed $\phi$ because temperature affects not only the dynamics of surfactants but also the kinetics of micellar scission and recombination, thereby altering the distribution of $N_\mathrm{ag}$.
As shown in a previous study,~\cite{Koide2023-yb} the mean number density of micelles increases with $k_BT$ at fixed $\phi$, corresponding to an increased population of small micelles.
This $k_BT$ dependence of the aggregation number distribution mainly contributes to the increase in $D/k_BT$ with $k_BT$.
In the following, we focus on $\phi$ dependence of $D$ in systems with $k_BT=1$ because the qualitative behavior is common for all values of $k_BT$ considered.

To reveal the physical mechanism underlying the nonmonotonic dependence of $D$ on $\phi$, we decompose MSD $\langle \Delta \bm{r}^2(t)\rangle$ of surfactants into two contributions: micellar center-of-mass diffusion and surfactant diffusion relative to the micellar center of mass.
Previous models~\cite{Jonstroemer1991-tq,Kato1993-yv,Kato1994-jt,Schmitt1998-pa} explained the nonmonotonicity of $D$ by employing the relation $D=D_\mathrm{m}+D_\mathrm{s}$, where $D_\mathrm{m}$ is the diffusion coefficient of micelles and $D_\mathrm{s}$ is the diffusion coefficient associated with surfactant diffusion within micelles.
In the present study, we directly evaluate the contributions of these two diffusion mechanisms in our systems.
Let us denote the MSD $\langle \Delta \bm{r}^2(t)\rangle$ of surfactants by $g(t)$.
We define the contribution $g_\mathrm{m}(t)$ of micellar center-of-mass diffusion to $g(t)$ as
\begin{equation}
    g_\mathrm{m}(t) = \left\langle\left|\int_0^t\bm{v}_\mathrm{m}({t^\prime})dt^\prime\right|^2\right\rangle. \label{eq:decomp_msd_m}
\end{equation}
Here, $\bm{v}_\mathrm{m}(t)$ is defined as the velocity of the micellar center of mass:
\begin{equation}
    \bm{v}_\mathrm{m}(t) = \frac{1}{N_\mathrm{ag}(t)}\sum_{j=1}^{N_\mathrm{ag}(t)} \bm{v}_j(t),
\end{equation}
where $N_\mathrm{ag}(t)$ is the aggregation number of the micelle at $t$ to which the target surfactant belongs, and $\bm{v}_j(t)$ denotes the velocity of the $j$-th surfactant belonging to the micelle.
Then, the contribution $g_\mathrm{s}(t)$ of surfactant diffusion relative to the micellar center of mass is defined as 
\begin{equation}
    g_\mathrm{s}(t) = g(t) - g_\mathrm{m}(t).\label{eq:decomp_msd_s}
\end{equation} 
Since micelles are likely to undergo scission and recombination over $t$, $g_\mathrm{m}(t)$ and $g_\mathrm{s}(t)$ inherently reflect the effects of these processes.
To numerically compute $g_\mathrm{m}(t)$, the trajectories of surfactants are sampled at discrete times $t_0=0, t_1=\delta t, \ldots, t_n = n\delta t(=t)$, where $\delta t$ is a fixed time interval.
We introduce the micellar displacement $\Delta \bm{r}_\mathrm{m}(t_{i+1}-t_i)$ within $\delta t$ as 
\begin{equation}
    \Delta \bm{r}_\mathrm{m}(t_{i+1}-t_i)= \frac{1}{N_\mathrm{ag}(t_i)}\sum_{j=1}^{N_\mathrm{ag}(t_i)}\Delta \bm{r}_j(t_{i+1}-t_i).\label{eq:mic_displacement}
\end{equation}
Here, $\Delta \bm{r}_j(t_{i+1}-t_i)$ denotes the displacement of the $j$-th surfactant belonging to the micelle.
It is worth emphasizing that although $\Delta \bm{r}_\mathrm{m}(t_{i+1}-t_i)$ does not correspond to the exact displacement of the micellar center of mass if scission or recombination occurs within $[t_i,t_{i+1}]$, we choose a sufficiently short time interval $\delta t=100\Delta t$ compared with characteristic timescales of micellar scission and recombination~(see Fig.~\ref{fig:timescale}).
By sequentially summing the displacement $\Delta \bm{r}_\mathrm{m} (t_{i+1}-t_i)$ within the short interval $\delta t$, we evaluate $g_\mathrm{m}(t)$ as
\begin{equation}
    g_\mathrm{m}(t) = \left\langle \left|\sum_{i=0}^{n-1} \Delta \bm{r}_\mathrm{m} (t_{i+1}-t_i)\right|^2\right\rangle.
\end{equation}

\begin{figure}
    \centering
        \begin{tabular}{c}
        \begin{minipage}{1\hsize}
            \begin{overpic}[width=1\linewidth]{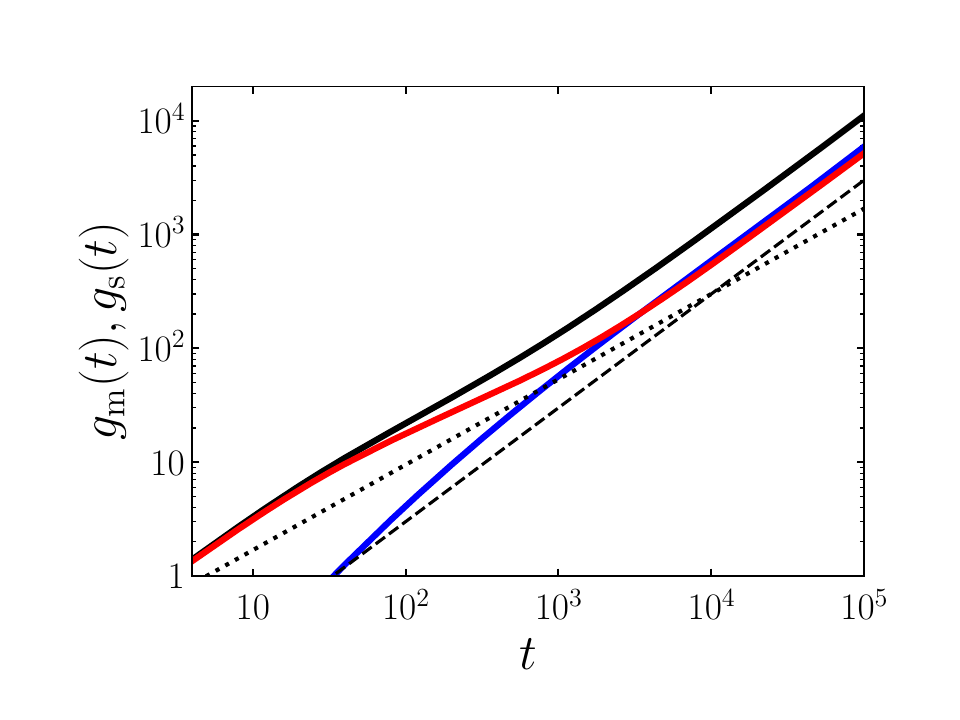}
                \linethickness{3pt}
          \put(5,63){(a)}

            \end{overpic}
        \end{minipage}\\
        \begin{minipage}{1\hsize}
            \begin{overpic}[width=1\linewidth]{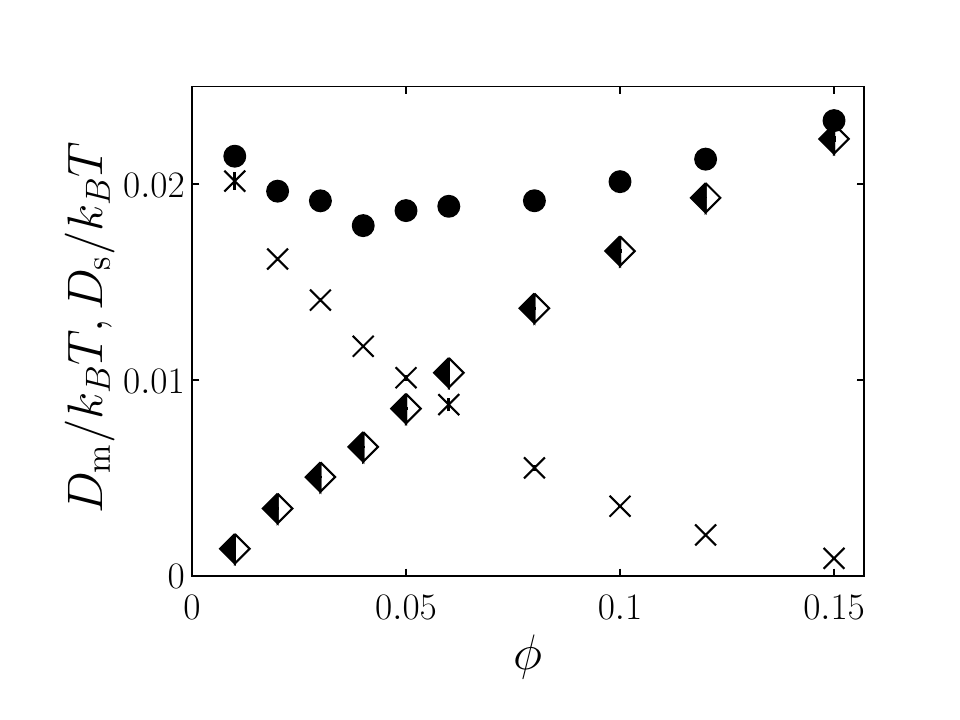}
                \linethickness{3pt}
          \put(5,63){(b)}

            \end{overpic}
        \end{minipage}
        \end{tabular}
        \caption{(a) Decomposed MSD of surfactants for $\phi=0.05$ with $k_BT=1$: blue solid line, $g_\mathrm{m}(t)$; red solid line, $g_\mathrm{s}(t)$. The black solid line indicates the MSD $g(t)(=\langle \Delta \bm{r}^2(t)\rangle)$ of surfactants~(the same as in Fig.~\ref{fig:msd}). The dotted and dashed lines indicate $g(t)\propto t^{0.75}$ and $g(t)\propto t$, respectively. (b) Decomposed self-diffusion coefficient of surfactants normalized by $k_BT$: cross, $D_\mathrm{m}/k_BT$; half-filled diamond, $D_\mathrm{s}/k_BT$. The filled circles indicate $D/k_BT$~(the same as in Fig.~\ref{fig:d_sur}). The error bars denote the standard deviations for three independent simulations.}
        \label{fig:decomp}
  \end{figure}

We demonstrate the competition between micellar center-of-mass diffusion and surfactant diffusion within micelles through the decomposed form $g(t)=g_\mathrm{m}(t)+g_\mathrm{s}(t)$ of the MSD.
Figure~\ref{fig:decomp}(a) shows $g_\mathrm{m}(t)$ and $g_\mathrm{s}(t)$ for $\phi=0.05$ with $k_BT=1$.
For comparison, we also present $g(t)$, as shown in Fig.~\ref{fig:msd}.
For $\phi=0.05$, $g_\mathrm{s}(t)$ dominates the subdiffusion at small $t$, whereas both $g_\mathrm{m}(t)$ and $g_\mathrm{s}(t)$ contribute to the normal diffusion at large $t$.
Again, it is worth emphasizing that $g_\mathrm{s}(t)$ represents the contribution of surfactant diffusion within micelles combined with scission and recombination processes.
We define the decomposed self-diffusion coefficients $D_\mathrm{m}$ and $D_\mathrm{s}$ for these two diffusion mechanisms by fitting $g_\mathrm{m}(t)$ and $g_\mathrm{s}(t)$ to $6D_\mathrm{m} t$ and $6D_\mathrm{s} t$, respectively.
Figure~\ref{fig:decomp}(b) shows $D_\mathrm{m}$ and $D_\mathrm{s}$ as functions of $\phi$ for $k_BT=1$.
Figure~\ref{fig:decomp}(b) demonstrates that while $D_\mathrm{m}$ is a monotonically decreasing function of $\phi$, $D_\mathrm{s}$ is a monotonically increasing function of $\phi$.
Consequently, as $\phi$ increases, the dominant contribution shifts from $D_\mathrm{m}$ to $D_\mathrm{s}$, resulting in the nonmonotonic dependence of $D$.
To summarize, decomposition analysis of MSD~[Eqs.~\eqref{eq:decomp_msd_m} and \eqref{eq:decomp_msd_s}] allows us to directly demonstrate that the nonmonotonic dependence of $D$ on $\phi$ results from the competition between micellar center-of-mass diffusion and surfactant diffusion within micelles.
This finding is consistent with the assumptions of previous models.~\cite{Jonstroemer1991-tq,Kato1993-yv,Kato1994-jt,Schmitt1998-pa}
We have also confirmed that the same crossover in the diffusion mechanism of surfactants occurs for all considered values of $k_BT$~(see Appendix B).

\section{\label{sec:discussion}Discussion}
In Sec.~\ref{sec:result}, we have revealed that the intriguing nonmonotonicity of $D$ arises from the competition between two diffusion mechanisms: micellar center-of-mass diffusion and surfactant diffusion within micelles.
This section aims to elucidate the origin of this competition from the perspective of the dynamics of individual micelles and surfactants and the kinetics of micellar scission and recombination.

\subsection{\label{subsec:micellar-center-of-mass-diffusion}Micellar center-of-mass diffusion}
\begin{figure}
    \centering
    \begin{overpic}[width=1\linewidth]{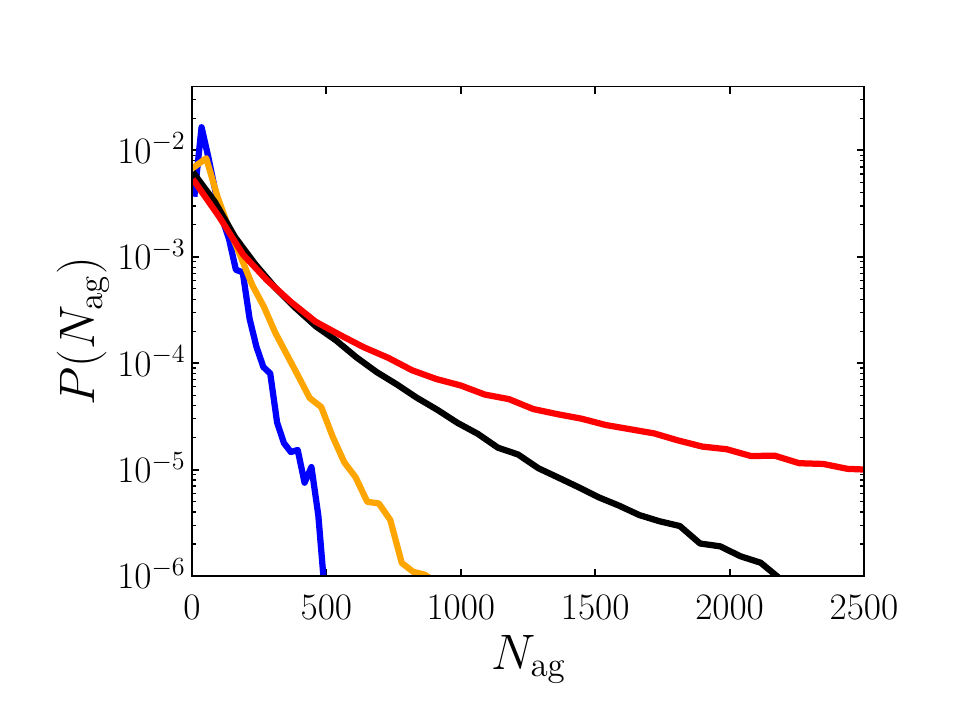} 
    \end{overpic}
    \caption{Probability density function $P(N_\mathrm{ag})$ of the aggregation number $N_\mathrm{ag}$ for $\phi=0.01$~(blue), $0.02$~(orange), $0.05$~(black), and $0.1$~(red) with $k_BT=1$.}
    \label{fig:nag_dis}
\end{figure}

This subsection focuses on the center-of-mass diffusion of individual micelles to investigate the decreasing tendency of $D$ with increasing $\phi$, which results from the monotonic decrease in the contribution $D_\mathrm{m}$ of micellar center-of-mass diffusion~[Fig.~\ref{fig:decomp}(b)].
As shown in Fig.~\ref{fig:snapshot}, surfactants form larger micelles at larger values of $\phi$.
Figure~\ref{fig:nag_dis} shows the probability density function $P(N_\mathrm{ag})$ of the aggregation number $N_\mathrm{ag}$ of micelles for various $\phi$ with $k_BT=1$.
We confirm that $P(N_\mathrm{ag})$ monotonically increases with $\phi$ for large $N_\mathrm{ag}$, indicating that there are many large micelles in systems with large $\phi$.

\begin{figure}
    \centering
    \begin{overpic}[width=1\linewidth]{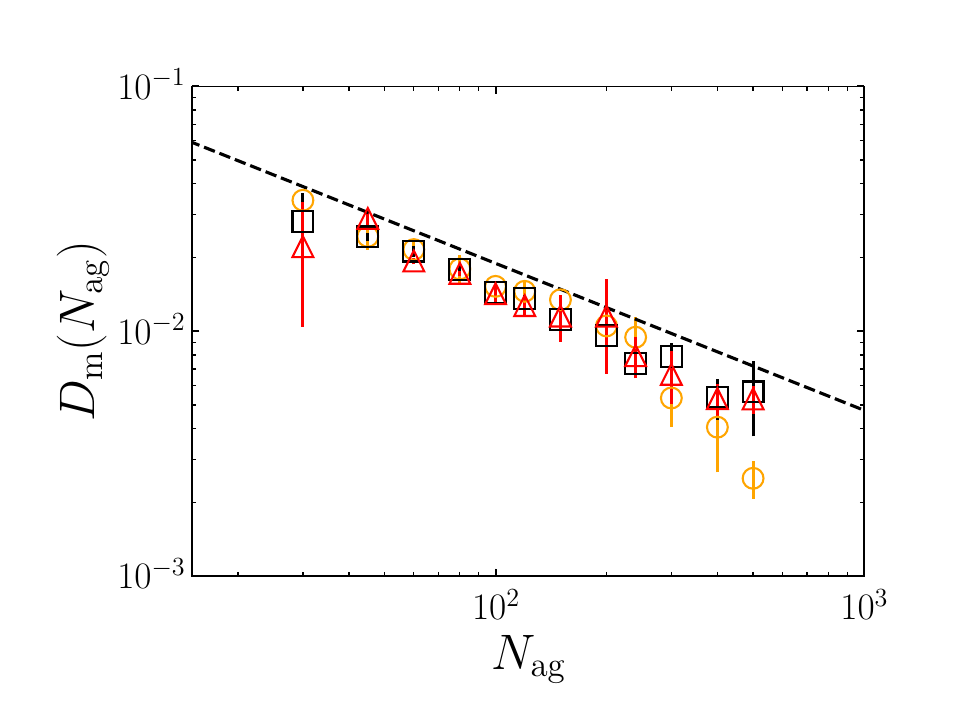} 
    \end{overpic}
    \caption{Diffusion coefficient $D_\mathrm{m}(N_\mathrm{ag})$ of the micellar center of mass as a function of the aggregation number $N_\mathrm{ag}$ for $\phi=0.02$~(orange), $0.05$~(black), and $0.1$~(red) with $k_BT=1$. The black dashed line indicates $D_\mathrm{m}(N_\mathrm{ag})\propto N_\mathrm{ag}^{-0.6}$. The error bars denote the standard deviations for three independent simulations.}
    \label{fig:d_mic}
\end{figure}

We investigate how the micellar diffusivity depends on $N_\mathrm{ag}$ and $\phi$.
Figure~\ref{fig:d_mic} shows the diffusion coefficient $D_\mathrm{m}(N_\mathrm{ag})$ of the center of mass of individual micelles as a function of $N_\mathrm{ag}$ for various $\phi$.
Since micelles with large $N_\mathrm{ag}$ have short lifetimes, as shown later~(see Fig.~\ref{fig:timescale}), it is difficult to evaluate their MSD at large $t$. 
To address this issue, we evaluate $D_\mathrm{m}(N_\mathrm{ag})$ as
\begin{equation}
    D_\mathrm{m}(N_\mathrm{ag})=\frac{1}{3}\int_0^\infty \langle \bm{v}_\mathrm{m}(t)\cdot \bm{v}_\mathrm{m}(0)\rangle_{N_\mathrm{ag}} dt,\label{eq:D_m}
\end{equation}
where $\bm{v}_\mathrm{m}(t)$ is the velocity of the center of mass of micelles and $\langle\cdot \rangle_{N_\mathrm{ag}}$ denotes the conditional ensemble average for fixed $N_\mathrm{ag}$.
Note that $\langle\cdot \rangle_{N_\mathrm{ag}}$ does not include the effect of scission and recombination processes.
To compute the conditional ensemble average for fixed $N_\mathrm{ag}$, we use data from micelles with aggregation numbers in the range $[N_\mathrm{ag}-\Delta N_\mathrm{ag}/2,N_\mathrm{ag}+\Delta N_\mathrm{ag}/2]$ until the micelles undergo scission or recombination.
Here, $\Delta N_\mathrm{ag}$ is set to $0.1N_\mathrm{ag}$, and the definition of micellar scission and recombination is the same as in previous studies.~\cite{Koide2022-bp,Koide2023-yb}
We define scission\,(recombination) as an event in which $N_\mathrm{ag}$ of a micelle decreases\,(increases) by more than the threshold value $n_\mathrm{mic}(=10)$ within a given time interval $\delta t(=100\Delta t)$.
Further details on the analysis of the micellar center of mass are provided in Appendix C.
Figure~\ref{fig:d_mic} demonstrates that $D_\mathrm{m}(N_\mathrm{ag})$ monotonically decreases as $N_\mathrm{ag}$ increases, indicating that larger micelles have lower diffusivity. 
We also find that $D_\mathrm{m}(N_\mathrm{ag})$ is almost independent of $\phi$ for fixed $N_\mathrm{ag}$, which indicates that inter-micellar interactions are not significant within the considered concentration range.
Therefore, the decrease in $D_\mathrm{m}$ with increasing $\phi$~[Fig.~\ref{fig:decomp}(b)] is explained by $\phi$ dependence of $P(N_\mathrm{ag})$~(Fig.~\ref{fig:nag_dis}) and $N_\mathrm{ag}$ dependence of $D_\mathrm{m}(N_\mathrm{ag})$~(Fig.~\ref{fig:d_mic}), because the number of micelles with large $N_\mathrm{ag}$ increases with $\phi$, and they have low $D_\mathrm{m}(N_\mathrm{ag})$.
Incidentally, $D_\mathrm{m}(N_\mathrm{ag})$ follows a power law $D_\mathrm{m}(N_\mathrm{ag})\propto N_\mathrm{ag}^{-0.6}$, which is close to the prediction of the Zimm model~\cite{Zimm1956-gt} and consistent with previous reports on hydrodynamic interactions in DPD simulations.~\cite{Jiang2007-rf}


\subsection{\label{subsec:relative-diffusion}Surfactant diffusion within micelles}

\begin{figure}
    \centering
    \begin{overpic}[width=1\linewidth]{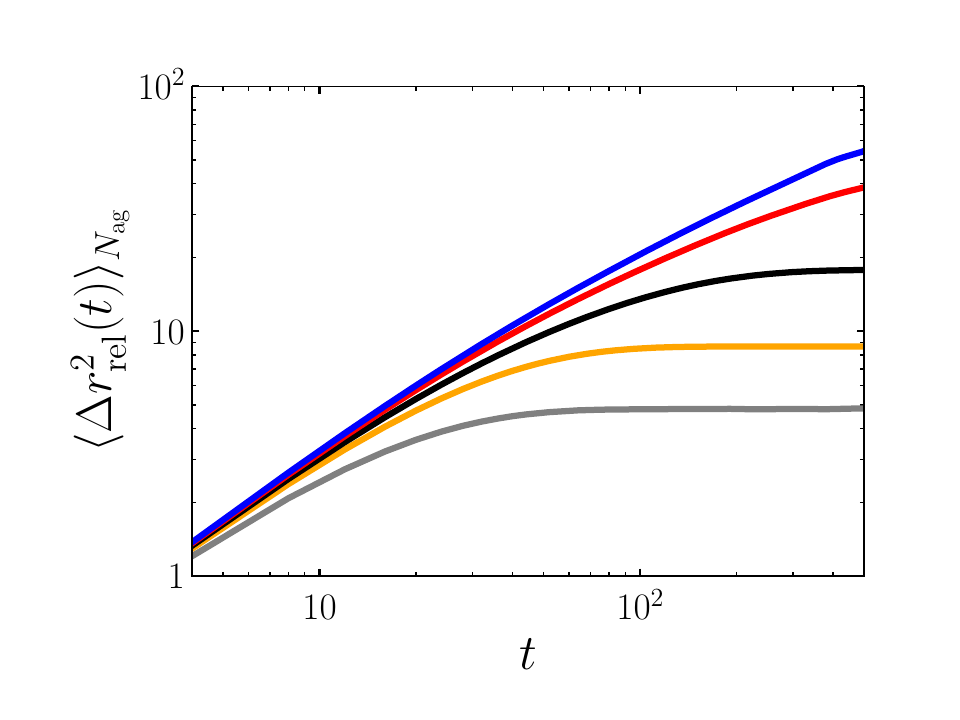} 
    \end{overpic}
    \caption{MSD $\langle \Delta \bm{r}_\mathrm{rel}^2(t)\rangle_{N_\mathrm{ag}}$ of surfactant molecules relative to the micellar center of mass for $N_\mathrm{ag}=30$~(gray), $60$~(orange), $100$~(black), $200$~(red), and $400$~(blue) with $k_BT=1$ and $\phi=0.05$.}
    \label{fig:msd_rela}
\end{figure}

In this subsection, we investigate the detailed behavior of individual surfactants when they belong to micelles.
To quantify the internal diffusion of surfactants within micelles, we evaluate MSD $\langle \Delta \bm{r}_\mathrm{rel}^2(t)\rangle_{N_\mathrm{ag}}$ of surfactants relative to the micellar center of mass, defined as
\begin{equation}
 \langle \Delta \bm{r}_\mathrm{rel}^2(t)\rangle_{N_\mathrm{ag}} =\langle |\bm{r}_\mathrm{rel}(t)-\bm{r}_\mathrm{rel}(0)|^2\rangle_{N_\mathrm{ag}},
\end{equation}
where $\bm{r}_\mathrm{rel}(t) = \bm{r}(t) - \bm{r}_\mathrm{m}(t)$ is the relative vector between $\bm{r}(t)$ and the center of mass $\bm{r}_\mathrm{m}(t)$ of the micelle to which the target surfactant belongs.
Note that we perform conditional ensemble average based on $N_\mathrm{ag}$ for $\langle \Delta \bm{r}_\mathrm{rel}^2(t)\rangle_{N_\mathrm{ag}}$, which can be regarded as an elementary process of surfactant diffusion within micelles.
Figure~\ref{fig:msd_rela} shows $\langle \Delta \bm{r}_\mathrm{rel}^2(t)\rangle_{N_\mathrm{ag}}$ for various $N_\mathrm{ag}$.
For a given $t$, $\langle \Delta \bm{r}_\mathrm{rel}^2(t)\rangle_{N_\mathrm{ag}}$ increases with $N_\mathrm{ag}$, indicating that surfactants in larger micelles exhibit a faster diffusion relative to the center of mass of micelles.
Since the population of large micelles increases with $\phi$~(Fig.~\ref{fig:nag_dis}), the enhanced relative diffusion with $N_\mathrm{ag}$ can contribute to the increase in $D_\mathrm{s}$ with $\phi$~[Fig.~\ref{fig:decomp}(b)].
It is worth emphasizing here that the kinetics of micellar scission and recombination is also a crucial factor influencing $D_\mathrm{s}$, as discussed later in Sec.~\ref{sec:discussion} C.

\begin{figure}
    \centering
    \begin{overpic}[width=1\linewidth]{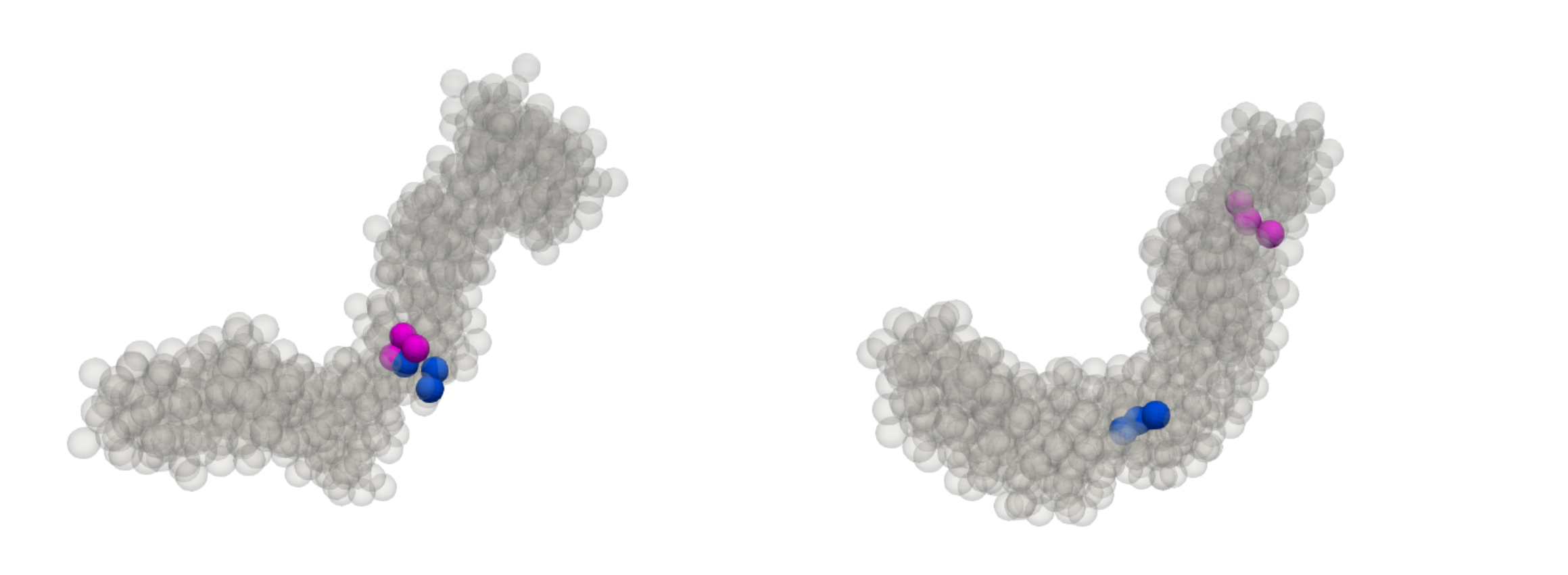} 
          \put(5,28){(a)}
          \put(55,28){(b)}

    \end{overpic}
    \caption{Visualization of two surfactant molecules~(magenta and blue) in a micelle~(white) which are initially located within $r_c$ at (a)~$t=0$ and move away from each other at (b)~$t=112$. }
    \label{fig:snap_two_monomer}
\end{figure}

\begin{figure}
    \centering
    \begin{overpic}[width=1\linewidth]{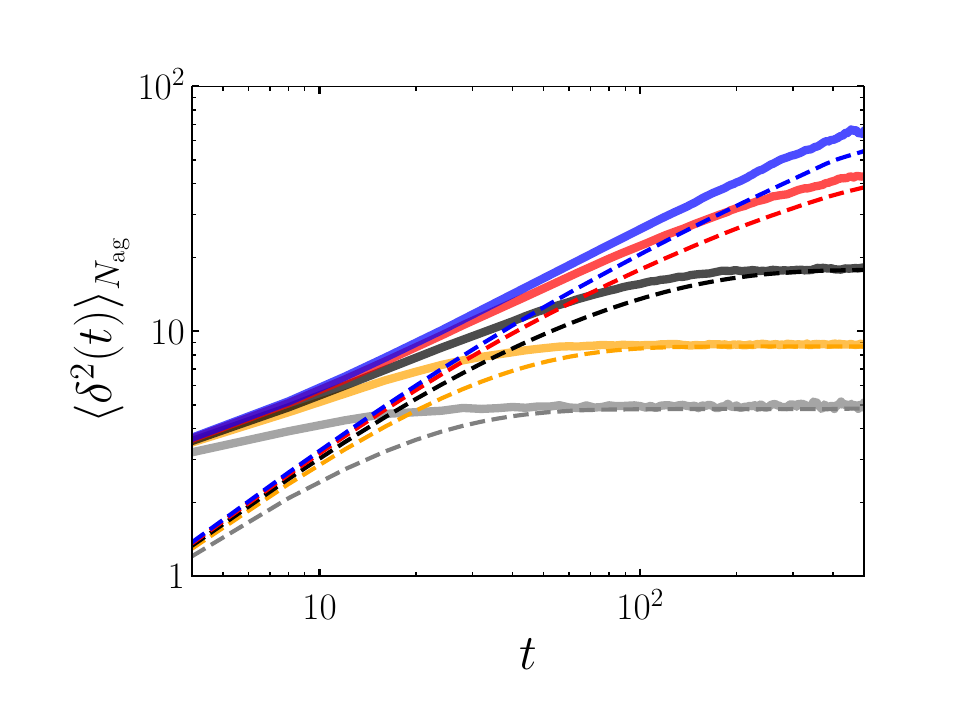} 
    \end{overpic}
    \caption{Mean-square distance $\langle\delta^2 (t)\rangle_{N_\mathrm{ag}}$ between two surfactant molecules in the same micelle with a fixed $N_\mathrm{ag}$ for $N_\mathrm{ag}=30$~(gray), $60$~(orange), $100$~(black), $200$~(red), and $400$~(blue) with $k_BT=1$ and $\phi=0.05$. The dashed lines indicate the MSD $\langle \Delta \bm{r}_\mathrm{rel}^2(t)\rangle_{N_\mathrm{ag}}$ of surfactant molecules relative to the center of mass of micelles (the same as in Fig.~\ref{fig:msd_rela}).}
    \label{fig:two_dis}
\end{figure}

We aim to identify the mechanism for the high relative diffusivity of surfactants in large micelles.
For this purpose, we focus on the distance $\delta(t)$ between two surfactants belonging to the same micelle, which directly characterizes the surfactant diffusion within micelles.
We monitor the time evolution of $\delta(t)$ of two surfactants whose centers of mass are within $r_c$ at the beginning, as shown in Fig.~\ref{fig:snap_two_monomer}. 
Figure~\ref{fig:two_dis} shows the ensemble average $\langle\delta^2(t)\rangle_{N_\mathrm{ag}}$ of $\delta^2(t)$ for each $N_\mathrm{ag}$.
We observe that $\langle\delta^2(t)\rangle_{N_\mathrm{ag}}$ increases with $t$, providing evidence that surfactants diffuse within micelles, which is a relevant process assumed in previous models to explain the increasing tendency of $D$.~\cite{Kato1993-yv,Kato1994-jt,Schmitt1998-pa}
For comparison, we also show $\langle \Delta \bm{r}_\mathrm{rel}^2(t)\rangle_{N_\mathrm{ag}}$ in Fig.~\ref{fig:two_dis}.
Similarly to $\langle \Delta \bm{r}_\mathrm{rel}^2(t)\rangle_{N_\mathrm{ag}}$ shown in Fig.~\ref{fig:msd_rela}, $\langle\delta^2(t)\rangle_{N_\mathrm{ag}}$ monotonically increases with $N_\mathrm{ag}$ for fixed $t$.
Finally, $\langle\delta^2(t)\rangle_{N_\mathrm{ag}}$ and $\langle \Delta \bm{r}_\mathrm{rel}^2(t)\rangle_{N_\mathrm{ag}}$ saturate to the same value because in the long-time limit $t\to \infty$, both quantities approach $2\langle \bm{r}_\mathrm{rel}^2 \rangle_{N_\mathrm{ag}}$.
The increase in $\langle\delta^2(t)\rangle_{N_\mathrm{ag}}$ with $N_\mathrm{ag}$ indicates that the distance between two surfactants in larger micelles grows larger.
Therefore, we conclude that the increase in $\langle \Delta \bm{r}_\mathrm{rel}^2(t)\rangle_{N_\mathrm{ag}}$ with $N_\mathrm{ag}$~(Fig.~\ref{fig:msd_rela}) is mainly due to the surfactant diffusion within micelles. 

\subsection{\label{subsec:timescale}Micellar scission and recombination}

Figure~\ref{fig:decomp} has demonstrated that the nonmonotonic behavior of $D$~(Fig.~\ref{fig:d_sur}) arises from the competition between micellar center-of-mass diffusion, which is a monotonically decreasing function of $N_\mathrm{ag}$~(Fig.~\ref{fig:d_mic}), and surfactant diffusion within micelles, which is a monotonically increasing function of $N_\mathrm{ag}$~(Fig.~\ref{fig:msd_rela}).
At small $\phi$, micellar center-of-mass diffusion is the dominant mechanism due to the prevalence of small micelles, whereas at large $\phi$, the dominant diffusion mechanism shifts to surfactant diffusion within large micelles.
However, without micellar kinetics, $D$ would be a monotonically decreasing function of $\phi$ because, in the long-time limit, surfactant diffusion is eventually dominated by micellar center-of-mass diffusion.
For surfactant diffusion within micelles to significantly contribute to the long-time surfactant diffusion, recombination has to occur on a relatively short timescale to facilitate the formation of diffusion paths.
In contrast, scission suppresses surfactant diffusion within micelles by shortening diffusion paths.
Thus, we have to consider the effect of micellar scission and recombination.
Previous models assumed scaling laws for the lifetime of micelles of average length~\cite{Morie1995-iq,Schmitt1998-pa} or the inter-micellar migration time of surfactants~\cite{Kato1993-yv,Kato1994-jt} to derive the concentration dependence of $D$.
In the present study, we directly evaluate the lifetime and recombination time of micelles for each $N_\mathrm{ag}$ to explore the relationship between micellar kinetics and surfactant diffusion.

\begin{figure}
    \centering
    \begin{overpic}[width=1\linewidth]{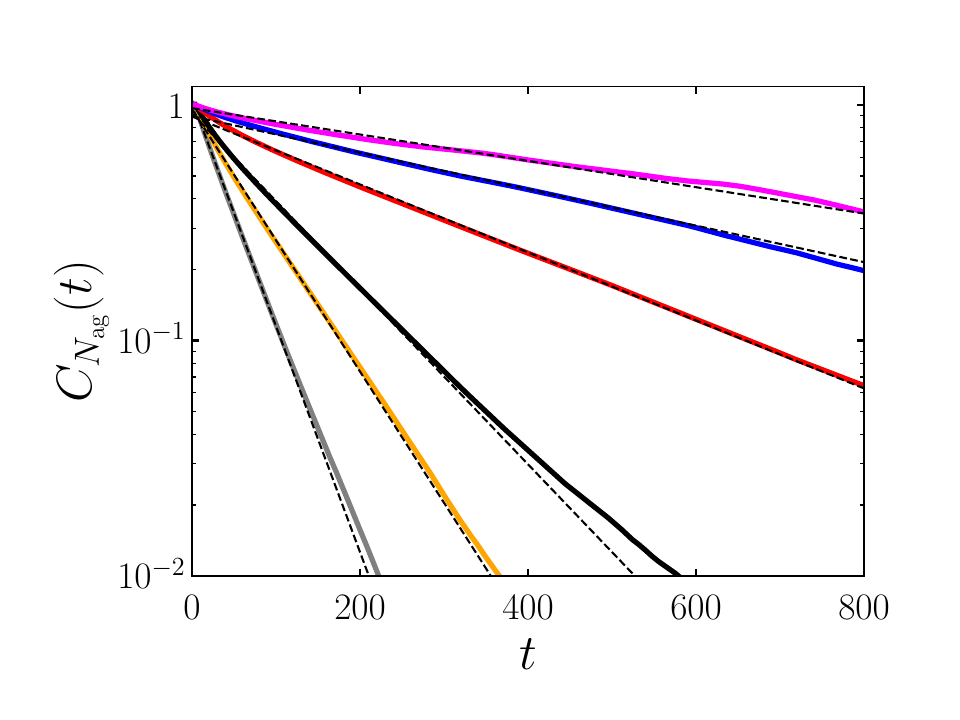} 
    \end{overpic}
    \caption{Autocorrelation function $C_{N_\mathrm{ag}}(t)$ of the relative vector $ \bm{r}_\mathrm{rel}(t)$ between the surfactant position $\bm{r}(t)$ and the center of mass $\bm{r}_\mathrm{m}(t)$ of the micelle for $k_BT=1$ and $\phi=0.04$ for $N_\mathrm{ag}=80$~(gray), $100$~(orange), $120$~(black), $200$~(red), $300$~(blue), and $400$~(magenta). The dashed lines indicate an exponential fit to $C_{N_\mathrm{ag}}(t)$.}
    \label{fig:auto_rela}
\end{figure}

We aim to understand the role of micellar scission and recombination in the nonmonotonicity of $D$ in terms of their timescales.
For this purpose, we focus on the competition among the average micellar lifetime $\tau_b(N_\mathrm{ag})$, the average micellar recombination time $\tau_r(N_\mathrm{ag})$, and the average internal diffusion time $\tau_d(N_\mathrm{ag})$ of surfactants within micelles.
We evaluate $\tau_b(N_\mathrm{ag})$ and $\tau_r(N_\mathrm{ag})$ using the same method as that employed in previous studies.~\cite{Koide2022-bp,Koide2023-yb}
To estimate $\tau_d(N_\mathrm{ag})$, we introduce the autocorrelation function $C_{N_\mathrm{ag}}(t)$ of the relative vector $ \bm{r}_\mathrm{rel}(t)$ between the surfactant position $\bm{r}(t)$ and the center of mass $\bm{r}_\mathrm{m}(t)$ of the micelle, defined as
\begin{equation}
    C_{N_\mathrm{ag}}(t) = \frac{\langle \bm{r}_\mathrm{rel}(t)\cdot \bm{r}_\mathrm{rel}(0)\rangle_{N_\mathrm{ag}}}{\langle \bm{r}_\mathrm{rel}^2\rangle_{N_\mathrm{ag}} }.
\end{equation}
Figure~\ref{fig:auto_rela} shows $C_{N_\mathrm{ag}}(t)$ for various $N_\mathrm{ag}$.
We find that $C_{N_\mathrm{ag}}(t)$ obeys an exponential function, similarly to the Ornstein--Uhlenbeck process.~\cite{uhlenbeck1930theory}
We evaluate $\tau_d(N_\mathrm{ag})$ by fitting $C_{N_\mathrm{ag}}(t)$ to $C_0\exp\{-t/\tau_d(N_\mathrm{ag})\}$.
Here, $\tau_d(N_\mathrm{ag})$ corresponds to the characteristic timescale of the surfactant diffusion within a given micelle with $N_\mathrm{ag}$.
Note that $\tau_d(N_\mathrm{ag})$ does not include the effect of scission and recombination at this point because $C_{N_\mathrm{ag}}(t)$ describes the relative diffusion of surfactants within micelles before the micelles undergo scission and recombination.

We begin by considering the competition between scission and internal diffusion while leaving recombination aside.
For $N_\mathrm{ag}$ such that $\tau_d(N_\mathrm{ag})\lesssim \tau_b(N_\mathrm{ag})$, since the lifetime of micelles is longer than the time required for surfactants to diffuse within micelles, surfactants can fully diffuse within micelles, and the effective internal diffusion time $\tau_d^\mathrm{eff}(N_\mathrm{ag})$ is equal to $\tau_d(N_\mathrm{ag})$.
In contrast, for $N_\mathrm{ag}$ such that $\tau_d(N_\mathrm{ag})\gtrsim \tau_b(N_\mathrm{ag})$, since micelles break before the internal diffusion has saturated, the surfactant diffusion within micelles is restricted by scission, and $\tau_d^\mathrm{eff}(N_\mathrm{ag})$ is bounded by $\tau_b(N_\mathrm{ag})$, i.e., $\tau_d^\mathrm{eff}(N_\mathrm{ag})\simeq \tau_b(N_\mathrm{ag})$.
Next, we consider the effect of recombination in addition to scission and internal diffusion.
For $\tau_r(N_\mathrm{ag})\gtrsim\tau_d^\mathrm{eff}(N_\mathrm{ag})$, the internal diffusion of surfactants saturates before recombination.
Consequently, the micellar center-of-mass diffusion mainly contributes to the surfactant diffusion.
In contrast, for $\tau_r(N_\mathrm{ag})\lesssim\tau_d^\mathrm{eff}(N_\mathrm{ag})$, micelles recombine with other micelles before the internal diffusion has saturated.
Together with this fast recombination, the contribution of the internal diffusion can be dominant over the micellar center-of-mass diffusion.
We propose that this competition among $\tau_b(N_\mathrm{ag})$, $\tau_r(N_\mathrm{ag})$, and $\tau_d(N_\mathrm{ag})$ is related to the dominant diffusion mechanism of surfactants.

\begin{figure}
    \centering
    \begin{overpic}[width=1\linewidth]{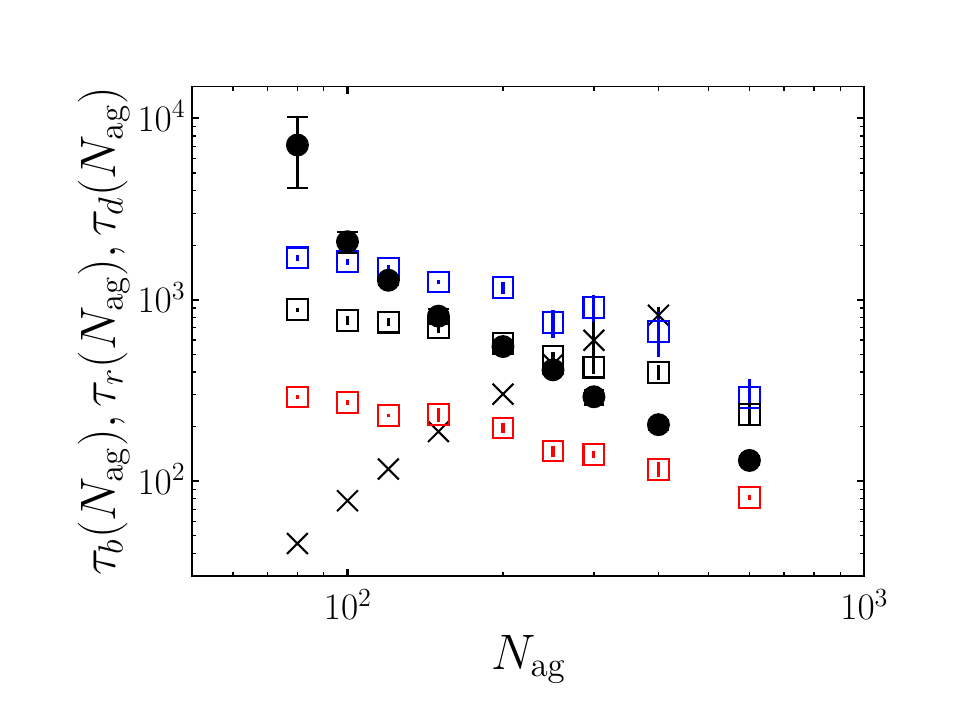} 
    \end{overpic}
    \caption{Average micellar lifetime $\tau_b(N_\mathrm{ag})$~(circle), average micellar recombination time $\tau_r(N_\mathrm{ag})$~(square), and average internal diffusion time $\tau_d(N_\mathrm{ag})$~(cross) of surfactants within micelles as functions of the aggregation number $N_\mathrm{ag}$ for $\phi=0.02$~(blue), $0.04$~(black), and $0.1$~(red). For clarity, $\tau_b(N_\mathrm{ag})$ and $\tau_d(N_\mathrm{ag})$ are shown only for $\phi=0.04$. The error bars denote the standard deviations for three independent simulations.}
    \label{fig:timescale}
\end{figure}

To verify this scenario, Fig.~\ref{fig:timescale} shows $\tau_b(N_\mathrm{ag})$, $\tau_r(N_\mathrm{ag})$, and $\tau_d(N_\mathrm{ag})$ with $k_BT=1$ for three values of $\phi$: $\phi =0.02(<\phi^*)$, $\phi=0.04(\simeq \phi^*)$, and $\phi=0.1(>\phi^*)$, where $D$ takes the minimum value at $\phi=\phi^*$. 
Note that we show $\tau_b(N_\mathrm{ag})$ and $\tau_d(N_\mathrm{ag})$ only for $\phi=0.04$ because $\tau_b(N_\mathrm{ag})$ and $\tau_d(N_\mathrm{ag})$ are almost independent of $\phi$ within the considered range~(see Appendix~D).
As reported in previous studies,~\cite{Koide2022-bp,Koide2023-yb} $\tau_b(N_\mathrm{ag})$ and $\tau_r(N_\mathrm{ag})$ are decreasing functions of $N_\mathrm{ag}$ and exhibit $\tau_b(N_\mathrm{ag})\propto N_\mathrm{ag}^{-1}$ and $\tau_r(N_\mathrm{ag})\propto N_\mathrm{ag}^{-1}$ for large $N_\mathrm{ag}$ due to the one-dimensional growth of wormlike micelles.
In addition, $N_\mathrm{ag}$ dependence of $\tau_r(N_\mathrm{ag})$ remains nearly unchanged for different values of $\phi$.
We observe that $\tau_d(N_\mathrm{ag})$ increases with $N_\mathrm{ag}$, indicating the longer internal diffusion of surfactants within larger micelles.
Since $\tau_b(N_\mathrm{ag})$ and $\tau_d(N_\mathrm{ag})$ intersect at $N_\mathrm{ag}\simeq 250$, $\tau_d^\mathrm{eff}(N_\mathrm{ag})\simeq \tau_d(N_\mathrm{ag})$ for $N_\mathrm{ag}\lesssim 250$, and $\tau_d^\mathrm{eff}(N_\mathrm{ag})\simeq \tau_b(N_\mathrm{ag})$ for $N_\mathrm{ag}\gtrsim 250$, as discussed above.
For $\phi=0.02$~(i.e., the decreasing region of $D$), $\tau_r(N_\mathrm{ag})\gtrsim\tau_d^\mathrm{eff}(N_\mathrm{ag})$ holds regardless of $N_\mathrm{ag}$. 
This relation indicates that the internal diffusion of surfactants within micelles is likely to saturate, and the contribution of micellar center-of-mass diffusion dominates the surfactant diffusion.
In contrast, for $\phi=0.04$~(i.e., the point of the minimum $D$), $\tau_r(N_\mathrm{ag})\simeq\tau_d^\mathrm{eff}(N_\mathrm{ag})$ for $N_\mathrm{ag}\simeq 250$ due to the decrease in $\tau_r(N_\mathrm{ag})$ with increasing $\phi$.
According to our scenario, the contribution of internal diffusion begins to emerge prominently.
For $\phi=0.1$~(i.e., the increasing region of $D$), the relation $\tau_r(N_\mathrm{ag})\lesssim\tau_d^\mathrm{eff}(N_\mathrm{ag})$ is satisfied for a broader range of $N_\mathrm{ag}$, indicating that the internal diffusion can be dominant over the micellar center-of-mass diffusion.
We have confirmed that the same relationship among $\tau_b(N_\mathrm{ag})$, $\tau_r(N_\mathrm{ag})$, and $\tau_d(N_\mathrm{ag})$ holds for other $k_BT$~(see Appendix E).
Thus, we demonstrate that in the region where $D$ increases, micellar recombination occurs on a relatively fast timescale compared with micellar scission and internal diffusion of surfactants.

\section{\label{sec:conclusion}Conclusion}

We have conducted DPD simulations of nonionic surfactant solutions for various surfactant volume fractions $\phi$ and temperatures $T$ to investigate the concentration dependence of the self-diffusion coefficient $D$ of surfactants.
One of the most important conclusions is that $D$ changes nonmonotonically with $\phi$~(Fig.~\ref{fig:d_sur}), as observed in previous experiments.~\cite{Nilsson1983-sb,Ott1990-pt,Khatory1993-fj,Kato1993-yv,Kato1994-jt,Morie1995-iq,Narayanan1997-bn,Narayanan1998-wi}
We have directly demonstrated that this nonmonotonicity of $D$ is caused by the competition between micellar center-of-mass diffusion and surfactant diffusion within micelles by decomposing $D$ into the corresponding contributions $D_\mathrm{m}$ and $D_\mathrm{s}$~(Fig.~\ref{fig:decomp}).
Thus, our simulation results have quantitatively verified the physical mechanism assumed in previous models.~\cite{Jonstroemer1991-tq,Kato1993-yv,Kato1994-jt,Schmitt1998-pa}

We have explained this competition between the two diffusion mechanisms by focusing on the aggregation number distribution, the diffusion behavior of individual surfactants and micelles, and the kinetics of micellar scission and recombination.
Regarding micellar diffusion, the diffusion coefficient $D_\mathrm{m}(N_\mathrm{ag})$ of the micellar center of mass is a monotonically decreasing function of $N_\mathrm{ag}$~(Fig.~\ref{fig:d_mic}), while the probability density function $P(N_\mathrm{ag})$ of $N_\mathrm{ag}$ has demonstrated that the population of large micelles increases with $\phi$~(Fig.~\ref{fig:nag_dis}).
Thus, micellar center-of-mass diffusion is responsible for the decrease in $D$.
To understand the subsequent increase in $D$ due to surfactant diffusion within micelles, we have evaluated the MSD $\langle \Delta \bm{r}_\mathrm{rel}(t)^2\rangle_{N_\mathrm{ag}}$ of the relative vector $\bm{r}_\mathrm{rel}(t)$ between the surfactant position $\bm{r}(t)$ and the center of mass $\bm{r}_\mathrm{m}(t)$ of the micelle for each $N_\mathrm{ag}$~(Fig.~\ref{fig:msd_rela}).
For fixed $t$, $\langle \Delta \bm{r}_\mathrm{rel}(t)^2\rangle_{N_\mathrm{ag}}$ increases with $N_\mathrm{ag}$, indicating that surfactant diffusion within micelles is promoted in large micelles.
The time evolution of the distance $\delta(t)$ between two surfactants located in the same micelle has provided evidence that surfactants indeed diffuse within micelles~(Figs.~\ref{fig:snap_two_monomer} and \ref{fig:two_dis}).
Thus, the increasing tendency of $D$ at large $\phi$ is based on the enhancement of surfactant diffusion within micelles due to the formation of large micelles.

We have also investigated the role of micellar scission and recombination in the competition between the two diffusion mechanisms, because surfactant diffusion within micelles cannot contribute to the long-time diffusion of surfactants without micellar kinetics.
We have revealed the competing effects of scission, recombination, and internal diffusion by evaluating their characteristic timescales for each $N_\mathrm{ag}$~(Figs.~\ref{fig:auto_rela} and \ref{fig:timescale}).
Scission suppresses the internal diffusion of surfactants by cutting their diffusion paths, whereas recombination enhances internal diffusion by generating new diffusion paths of surfactants within micelles.
We have demonstrated that in the region where $D$ increases, micellar recombination occurs on a relatively fast timescale compared with micellar scission and internal diffusion of surfactants.

Although similar physical mechanisms of the nonmonotonic behavior of $D$ have been proposed in previous studies~\cite{Jonstroemer1991-tq,Kato1993-yv,Kato1994-jt,Morie1995-iq,Schmitt1998-pa} based on experiments and phenomenological models, this study has demonstrated the nonmonotonic behavior of $D$ using molecular simulations and explained its origin in terms of three factors that are challenging to measure simultaneously in experiments: the aggregation number distribution, the dynamics of individual surfactants and micelles, and the kinetics of micellar scission and recombination.
Specifically, at small $\phi$, surfactant diffusion within micelles has a negligible effect due to the prevalence of small micelles and the long recombination times in dilute solutions.
In this regime, the micellar center-of-mass diffusion is the dominant mechanism, leading to a decrease in $D$ with increasing $\phi$.
In contrast, at large $\phi$, surfactant diffusion within micelles, combined with fast recombination, dominantly contributes to the long-time diffusion of surfactants.
As $\phi$ increases, this diffusion mechanism is enhanced by promoting recombination, resulting in an increase in $D$.
\begin{acknowledgments}
    This work was supported by JSPS Grants-in-Aid for Scientific Research (21J21061 and 24KJ0109) and JST, ACT-X (JPMJAX24D5). 
    The DPD simulations were mainly conducted under the auspices of the NIFS Collaboration Research Programs (NIFS22KISS010).
    A part of the simulations was conducted using the JAXA Supercomputer System Generation 3 (JSS3).
\end{acknowledgments}
\section*{\label{sec:Appendix_A}Appendix A: System-size dependence of the self-diffusion coefficient}
\begin{figure}
    \centering
    \begin{overpic}[width=1\linewidth]{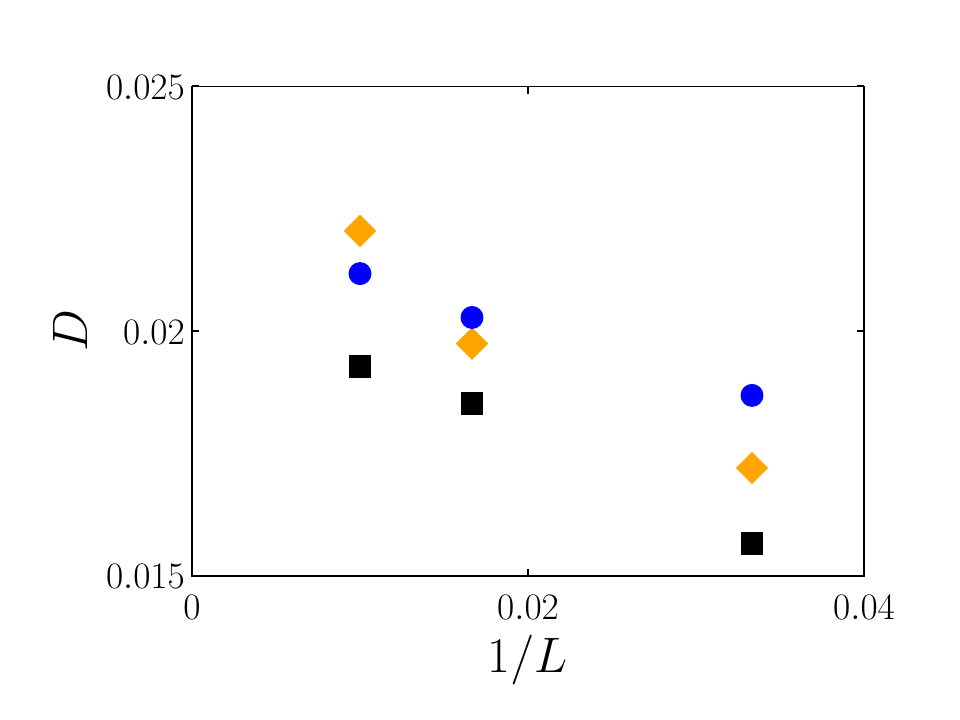} 
    \end{overpic}
    \caption{Self-diffusion coefficient $D$ of surfactants as a function of the inverse box length $1/L$ for $\phi=0.02$~(orange diamond), $0.05$~(black square), and $0.1$~(blue circle).}
    \label{fig:d_system}
\end{figure}

\renewcommand{\theequation}{A\arabic{equation} }
This Appendix demonstrates that the nonmonotonic concentration dependence of the self-diffusion coefficient $D$ of surfactants holds for different system sizes.
Figure~\ref{fig:d_system} shows $D$ as a function of the inverse box length $1/L$ for various surfactant volume fractions $\phi$.
Although, as is well known, $D$ strongly depends on $1/L$,~\cite{Yeh2004-lk} we confirm that $D$ for $\phi=0.05$ is smaller than that for $\phi=0.02$ and $0.1$, regardless of $L$.
Thus, we conclude that the nonmonotonic concentration dependence of $D$ is not attributed to the finite-size effect.
\section*{\label{sec:Appendix_B}Appendix B: Decomposition of the diffusion coefficient for different temperatures}
\begin{figure*}
    \centering
    \begin{overpic}[width=1\linewidth]{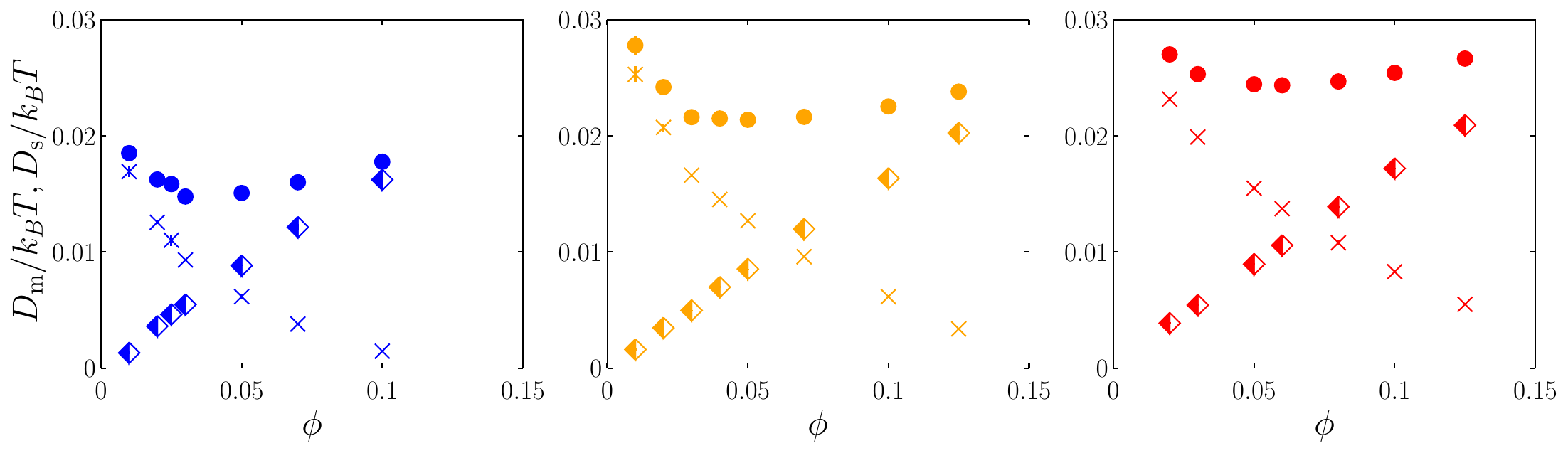} 
        \put(1,30){(a)}
        \put(34,30){(b)}
        \put(67,30){(c)}
    \end{overpic}
    \caption{Decomposed self-diffusion coefficient of surfactants normalized by $k_BT$ for (a) $k_BT=0.9$, (b) $1.1$, and (c) $1.2$: cross, $D_\mathrm{m}/k_BT$; half-filled diamond, $D_\mathrm{s}/k_BT$. The filled circles indicate $D/k_BT$~(the same as in Fig.~\ref{fig:d_sur}). The error bars denote the standard deviations for three independent simulations.}
    \label{fig:decomp_kbt}
\end{figure*}

\renewcommand{\theequation}{B\arabic{equation} }
In the main text, we demonstrate that for $k_BT=1$, the decomposition of $D$ into two contributions $D_\mathrm{m}$ and $D_\mathrm{s}$ explains the origin of the nonmonotonicity of $D$.
Figure~\ref{fig:decomp_kbt} shows $D_\mathrm{m}$ and $D_\mathrm{s}$ normalized by $k_BT$ as functions of $\phi$ for $k_BT=0.9$, $1.1$, and $1.2$.
For comparison, we also present $D/k_BT$ for each $k_BT$ in Fig.~\ref{fig:decomp_kbt}.
For all considered values of $k_BT$, the dominant diffusion mechanism shifts from micellar center-of-mass diffusion to surfactant diffusion within micelles as $\phi$ increases.

\section*{\label{sec:Appendix_C}Appendix C: Diffusion properties of the micellar center of mass}
\begin{figure}
    \centering
        \begin{tabular}{c}
        \begin{minipage}{1\hsize}
            \begin{overpic}[width=1\linewidth]{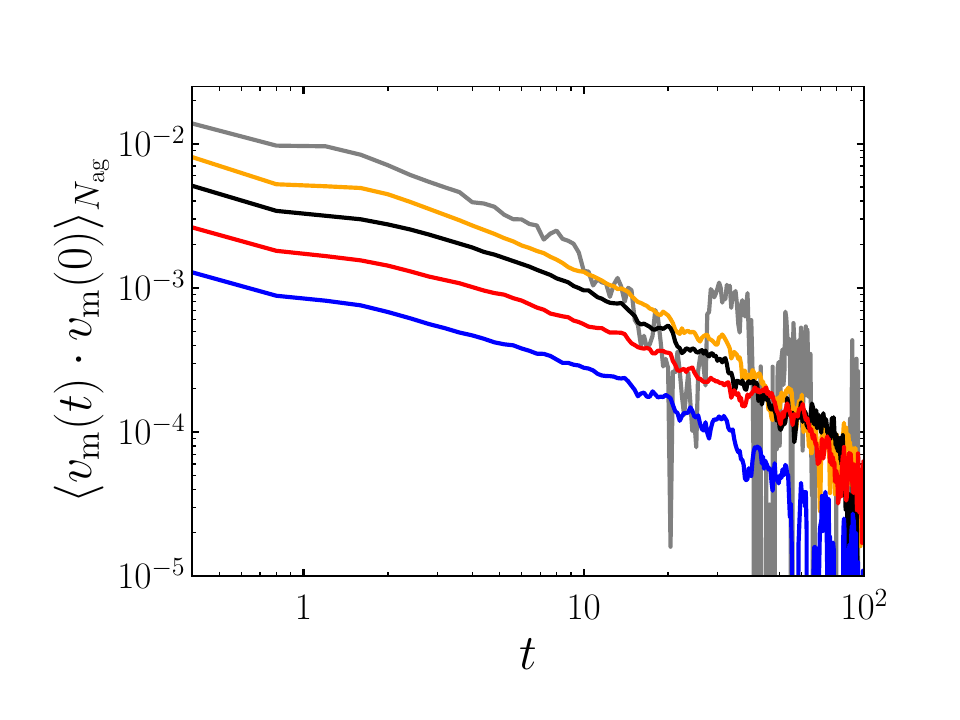}
                \linethickness{3pt}
          \put(5,63){(a)}

            \end{overpic}
        \end{minipage}\\
        \begin{minipage}{1\hsize}
            \begin{overpic}[width=1\linewidth]{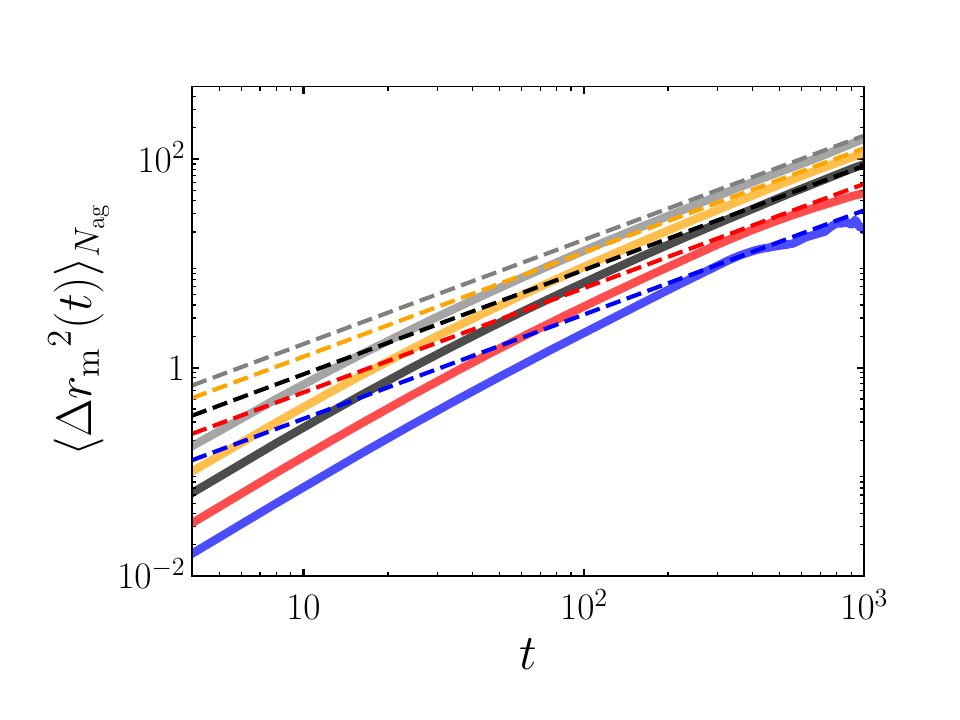}
                \linethickness{3pt}
          \put(5,63){(b)}

            \end{overpic}
        \end{minipage}

        \end{tabular}
  
        \caption{(a) Velocity autocorrelation function $\langle \bm{v}_\mathrm{m}(t)\cdot \bm{v}_\mathrm{m}(0)\rangle_{N_\mathrm{ag}}$ and (b) MSD $\langle \Delta {\bm{r}_\mathrm{m}}^2(t)\rangle_{N_\mathrm{ag}}$ of the micellar center of mass for $N_\mathrm{ag}=30$~(gray), $60$~(orange), $100$~(black), $200$~(red), and $400$~(blue) with $k_BT=1$ and $\phi=0.05$. In (b), the dashed lines indicate $6D_\mathrm{m}(N_\mathrm{ag})t$, where $D_\mathrm{m}(N_\mathrm{ag})$ is obtained using Eq.~\eqref{eq:D_m}.}
        \label{fig:micelle_com_validation}
  \end{figure}

\renewcommand{\theequation}{C\arabic{equation} }
In this Appendix, we provide detailed information on the analysis of the diffusion properties of the micellar center of mass.
In the main text, we estimate the diffusion coefficient $D_\mathrm{m}(N_\mathrm{ag})$ of the micellar center of mass using its velocity autocorrelation function $\langle \bm{v}_\mathrm{m}(t)\cdot \bm{v}_\mathrm{m}(0)\rangle_{N_\mathrm{ag}}$~[Eq.~\eqref{eq:D_m}].
Figure~\ref{fig:micelle_com_validation}(a) shows $\langle \bm{v}_\mathrm{m}(t)\cdot \bm{v}_\mathrm{m}(0)\rangle_{N_\mathrm{ag}}$ for various values of $N_\mathrm{ag}$.
Since micelles undergo scission and recombination, it is challenging to obtain accurate values of $\langle \bm{v}_\mathrm{m}(t)\cdot \bm{v}_\mathrm{m}(0)\rangle_{N_\mathrm{ag}}$ at large $t$.
Although the contribution from the tail region of $\langle \bm{v}_\mathrm{m}(t)\cdot \bm{v}_\mathrm{m}(0)\rangle_{N_\mathrm{ag}}$ cannot be precisely incorporated, Eq.~\eqref{eq:D_m} allows us to estimate $D_\mathrm{m}(N_\mathrm{ag})$ by integrating $\langle \bm{v}_\mathrm{m}(t)\cdot \bm{v}_\mathrm{m}(0)\rangle_{N_\mathrm{ag}}$.
Indeed, Fig.~\ref{fig:micelle_com_validation}(b) demonstrates that the MSD $\langle \Delta {\bm{r}_\mathrm{m}}^2(t)\rangle_{N_\mathrm{ag}}$ of the micellar center of mass approaches to $6D_\mathrm{m}(N_\mathrm{ag})t$ at large $t$, where $D_\mathrm{m}(N_\mathrm{ag})$ is obtained using Eq.~\eqref{eq:D_m}.
Thus, we confirm the reliability of our results based on Eq.~\eqref{eq:D_m}.
\section*{\label{sec:Appendix_D}Appendix D: Concentration dependence of diffusion and kinetics}
\begin{figure}
    \centering
        \begin{tabular}{c}
        \begin{minipage}{1\hsize}
            \begin{overpic}[width=1\linewidth]{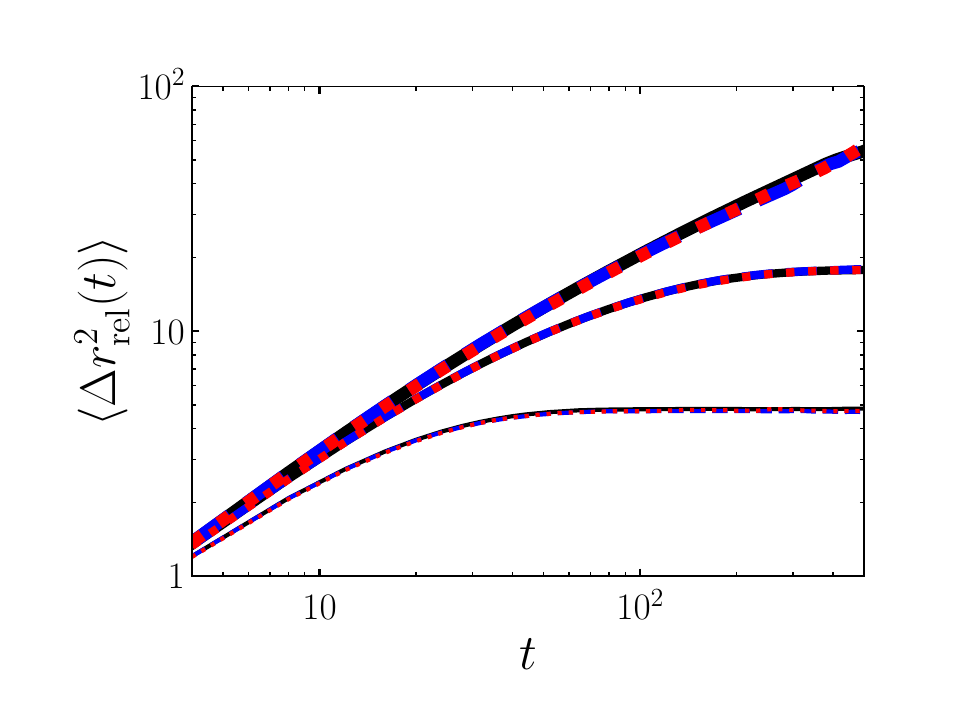}
                \linethickness{3pt}
          \put(5,63){(a)}

            \end{overpic}
        \end{minipage}\\
        \begin{minipage}{1\hsize}
            \begin{overpic}[width=1\linewidth]{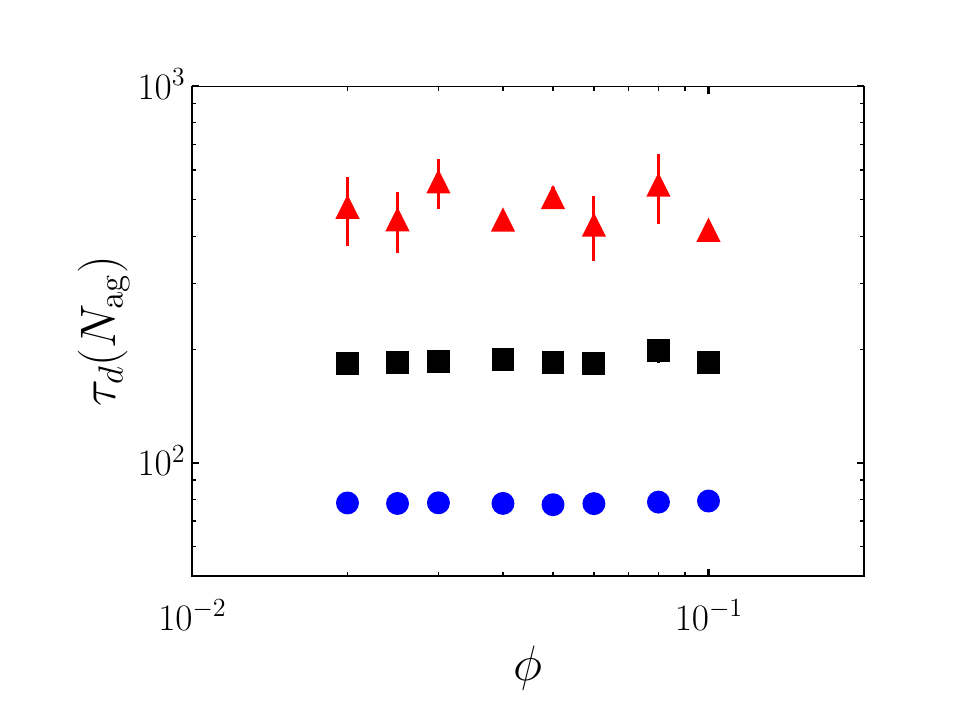}
                \linethickness{3pt}
          \put(5,63){(b)}

            \end{overpic}
        \end{minipage}

        \end{tabular}
  
        \caption{(a) MSD $\langle \Delta \bm{r}_\mathrm{rel}^2(t)\rangle_{N_\mathrm{ag}}$ of surfactant molecules relative to the micellar center of mass for $\phi=0.02$\,(blue dashed line), $0.05$\,(black solid line), and $0.1$\,(red dotted line) with $k_BT=1$. From thinner to thicker lines, $N_\mathrm{ag}=30$, $100$, and $400$. (b) Average internal diffusion time $\tau_d(N_\mathrm{ag})$ as a function of the surfactant volume fraction $\phi$ for $N_\mathrm{ag}=100$~(blue circle), $150$~(black square), and $250$~(red triangle). The error bars denote the standard deviations for three independent simulations.}
        \label{fig:internal_diffusion_phi}
  \end{figure}

\renewcommand{\theequation}{D\arabic{equation} }
In this Appendix, we demonstrate the concentration dependence of diffusion and kinetics.
We have already confirmed that the diffusion coefficient $D_\mathrm{m}(N_\mathrm{ag})$ of the center of mass of micelles is independent of $\phi$ within the considered range~(Fig.~\ref{fig:d_mic}).
Figure~\ref{fig:internal_diffusion_phi}(a) shows the MSD $\langle \Delta \bm{r}_\mathrm{rel}^2(t)\rangle_{N_\mathrm{ag}}$ of surfactants relative to the micellar center of mass for various values of $N_\mathrm{ag}$ and $\phi$.
We confirm that $\langle \Delta \bm{r}_\mathrm{rel}^2(t)\rangle_{N_\mathrm{ag}}$ increases with $N_\mathrm{ag}$ for fixed $t$, and $\langle \Delta \bm{r}_\mathrm{rel}^2(t)\rangle_{N_\mathrm{ag}}$ is almost insensitive to $\phi$.
Figure~\ref{fig:internal_diffusion_phi}(b) presents the average internal diffusion time $\tau_d(N_\mathrm{ag})$ as a function of $\phi$ for various $N_\mathrm{ag}$.
We confirm that $\tau_d(N_\mathrm{ag})$ is almost independent of $\phi$.
Thus, the concentration does not significantly alter the diffusivity of surfactants within micelles with fixed $N_\mathrm{ag}$ within the considered range.

\begin{figure}
    \centering
        \begin{tabular}{c}
        \begin{minipage}{1\hsize}
            \begin{overpic}[width=1\linewidth]{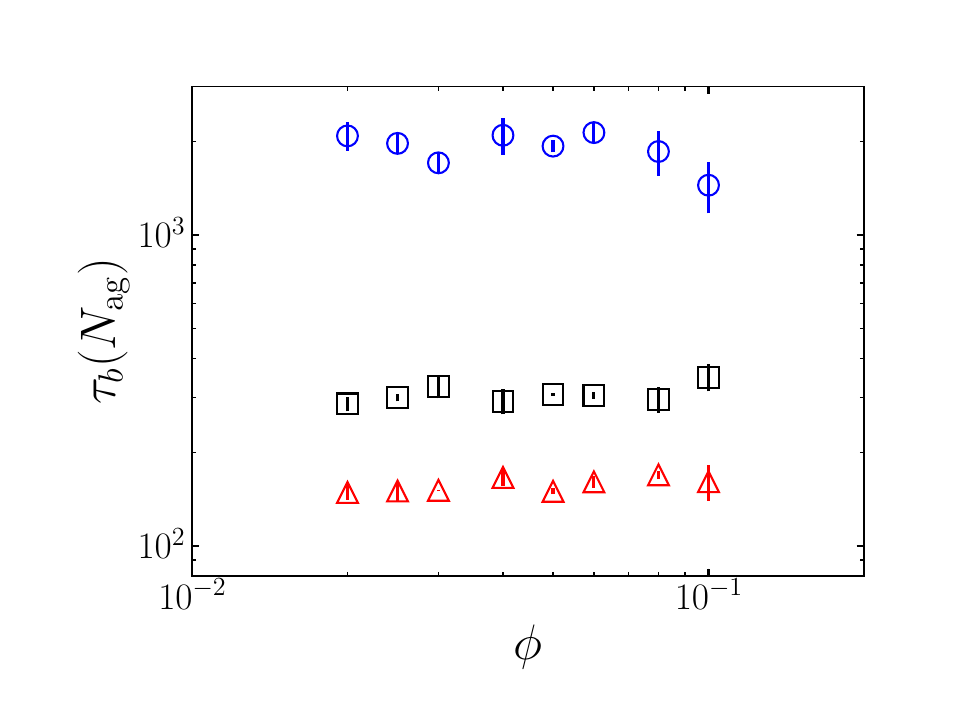}
                \linethickness{3pt}
          \put(5,63){(a)}

            \end{overpic}
        \end{minipage}\\
        \begin{minipage}{1\hsize}
            \begin{overpic}[width=1\linewidth]{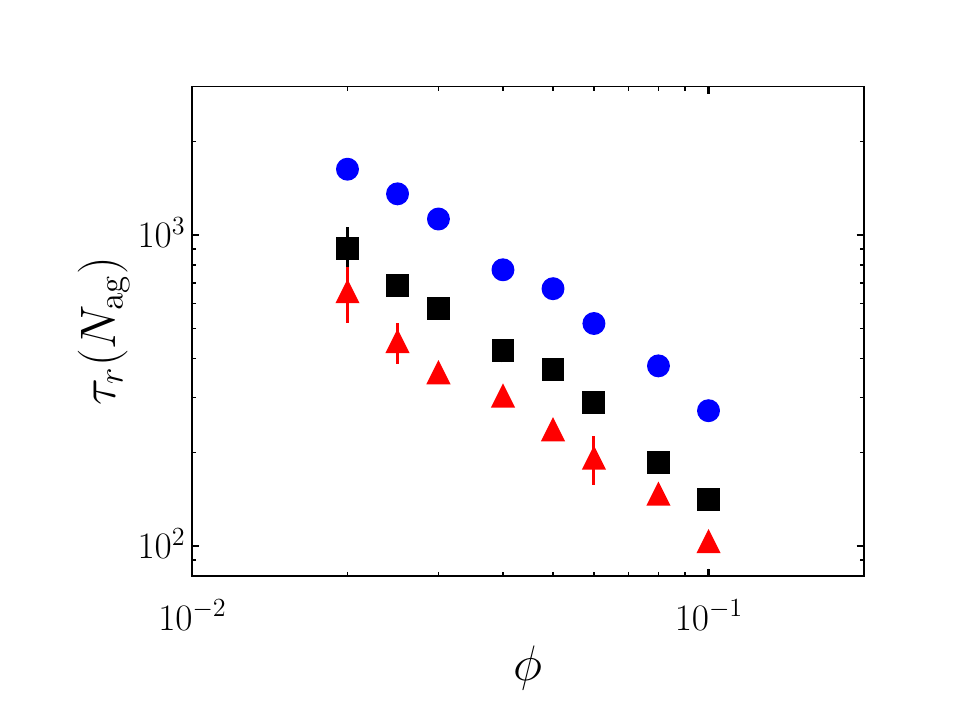}
                \linethickness{3pt}
          \put(5,63){(b)}

            \end{overpic}
        \end{minipage}

        \end{tabular}
  
        \caption{(a) Average micellar lifetime $\tau_b(N_\mathrm{ag})$ and (b) average micellar recombination time $\tau_r(N_\mathrm{ag})$ as functions of the surfactant volume fraction $\phi$ for $N_\mathrm{ag}=100$~(blue circle), $300$~(black square), and $500$~(red triangle). The error bars denote the standard deviations for three independent simulations.}
        \label{fig:kinetics_phi}
  \end{figure}

Figure~\ref{fig:kinetics_phi} shows the average lifetime $\tau_b(N_\mathrm{ag})$ and the average recombination time $\tau_r(N_\mathrm{ag})$ of micelles for various $N_\mathrm{ag}$ as functions of $\phi$.
While $\tau_b(N_\mathrm{ag})$ is independent of $\phi$, $\tau_r(N_\mathrm{ag})$ strongly depends on $\phi$, as reported in a previous study.~\cite{Koide2023-yb} 
Thus, in the main text, we focus on the concentration dependence of $\tau_r(N_\mathrm{ag})$ to explain the competition among scission, recombination, and internal diffusion~(Fig.\ref{fig:timescale}).

\section*{\label{sec:Appendix_E}Appendix E: Temperature dependence of the timescales of diffusion and kinetics}
\begin{figure*}
    \centering
    \begin{overpic}[width=1\linewidth]{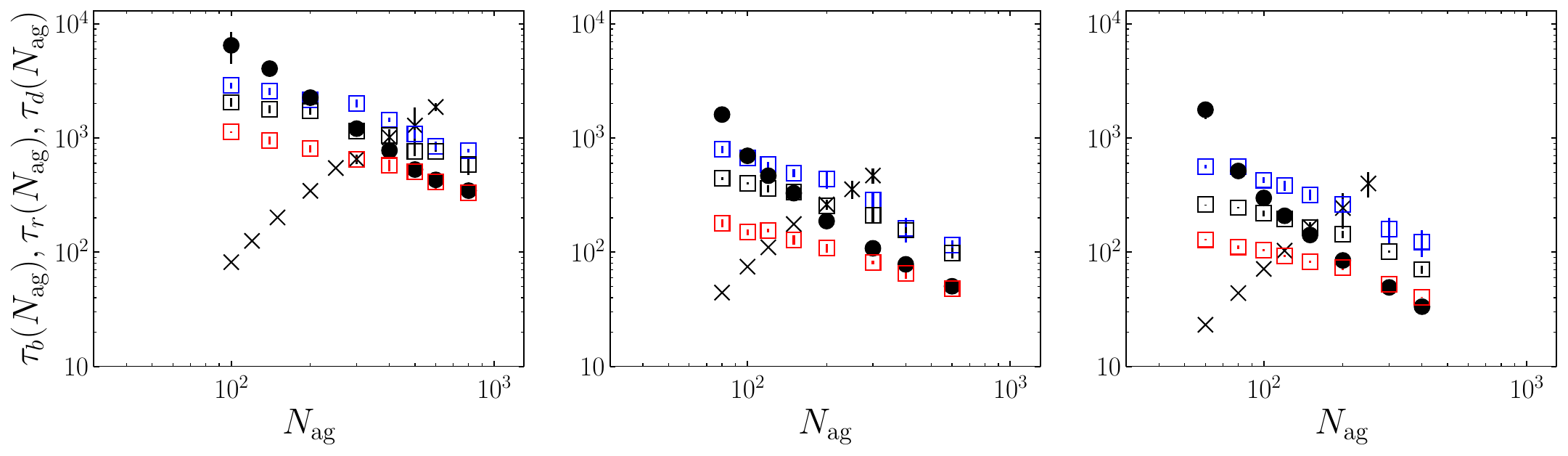} 
        \put(1,30){(a)}
        \put(34,30){(b)}
        \put(67,30){(c)}

    \end{overpic}
    \caption{Average micellar lifetime $\tau_b(N_\mathrm{ag})$~(circle), average micellar recombination time $\tau_r(N_\mathrm{ag})$~(square), and average internal diffusion time $\tau_d(N_\mathrm{ag})$~(cross) of surfactants within micelles as functions of the aggregation number $N_\mathrm{ag}$. In (a), $\phi=0.02$~(blue), $0.03$~(black), and $0.05$~(red) with $k_BT=0.9$. In (b), $\phi=0.03$~(blue), $0.05$~(black), and $0.1$~(red) with $k_BT=1.1$. In (c), $\phi=0.03$~(blue), $0.06$~(black), and $0.1$~(red) with $k_BT=1.2$. For clarity, $\tau_b(N_\mathrm{ag})$ and $\tau_d(N_\mathrm{ag})$ are shown only for a certain $\phi$. The error bars denote the standard deviations for three independent simulations.}
    \label{fig:dif_kine_temp}
\end{figure*}

\renewcommand{\theequation}{E\arabic{equation} }
In the main text, we have explained the competition among the timescales of scission, recombination, and internal diffusion for $k_BT=1$~(Fig.~\ref{fig:timescale}).
Here, we validate our scenario described in Sec.~\ref{subsec:timescale} by showing the results for different temperatures.
Figure~\ref{fig:dif_kine_temp} shows the average lifetime $\tau_b(N_\mathrm{ag})$, the average recombination time $\tau_r(N_\mathrm{ag})$, and the average internal diffusion time $\tau_d(N_\mathrm{ag})$ for $k_BT=0.9$, $1.1$, and $1.2$.
For each value of $k_BT$, we choose three values of $\phi$: $\phi <\phi^*$, $\phi\simeq \phi^*$, and $\phi>\phi^*$, where $D$ takes the minimum value at $\phi=\phi^*$.
As shown in Fig.~\ref{fig:timescale} for $k_BT=1$, $\tau_r(N_\mathrm{ag})$ decreases with increasing $\phi$ and almost intersects with both $\tau_b(N_\mathrm{ag})$ and $\tau_d(N_\mathrm{ag})$ at a certain $N_\mathrm{ag}$ for $\phi\simeq \phi^*$, indicating that the scenario proposed in Sec.~\ref{subsec:timescale} holds for all considered values of $k_BT$.

\setcounter{equation}{0}

    \section*{Author Declarations}
    \section*{Conflict of Interest}
    The author has no conflicts to disclose.
    \section*{Data Availability}
    The data that support the findings of
    this study are available from the
    corresponding author upon reasonable
    request.

\end{document}